\documentclass[]{iopart}

\usepackage{graphicx,amssymb}
\usepackage[usenames,dvipsnames]{xcolor}
\usepackage[normalem]{ulem} 
\usepackage{lscape}
\expandafter\let\csname equation*\endcsname\relax
\expandafter\let\csname endequation*\endcsname\relax
\usepackage{amsmath}
\usepackage[retainorgcmds]{IEEEtrantools}
\usepackage{cite} 
\usepackage[symbol*]{footmisc}
\usepackage{url} 
\usepackage{hyperref}

\definecolor{darkgreen}{rgb}{0.2,0.7,0.2}

\newcommand{\vS}{\mbox{\boldmath${S}$}}

\newcommand{\psifour}{\Psi_4}

\newcommand{\scriplus}{\mathcal{J}^+}
\newcommand{\rext}{\mathcal{R}}
\newcommand{\tjunk}{t_0}
\newcommand{\reftime}{t_{\mathrm{ref}}} 
\newcommand{\omref}{\omega_{22, \text{ ref}}}

\def\laq{\raise 0.4ex\hbox{$<$}\kern -0.8em\lower 0.62ex\hbox{$\sim$}}
\def\gaq{\raise 0.4ex\hbox{$>$}\kern -0.7em\lower 0.62ex\hbox{$\sim$}}

\sloppypar

\begin{document}

\title[Error-analysis and comparison to analytical models of 
numerical waveforms ...]{Error-analysis and comparison 
to analytical models of numerical waveforms produced by the NRAR Collaboration}

\author{Ian Hinder$^1$,
        Alessandra Buonanno$^2$,
        Michael Boyle$^3$,
        Zachariah~B. Etienne$^4$,
        James Healy$^5$,
        Nathan~K. Johnson-McDaniel$^6$,
        Alessandro Nagar$^7$,
        Hiroyuki Nakano$^{8,9}$,
        Yi Pan$^2$,
        Harald~P. Pfeiffer$^{10,11}$,
        Michael P{\"u}rrer$^{12}$,
        Christian Reisswig$^{13}$,
        Mark~A. Scheel$^{13}$,
        Erik Schnetter$^{14,15,16}$,
        Ulrich Sperhake$^{13,17,18,19}$,
        Bela Szil\'{a}gyi$^{13}$,
        Wolfgang Tichy$^{20}$,
        Barry Wardell$^{1,21}$,
        An{\i}l Zengino\u{g}lu$^{13}$,
        Daniela Alic$^{1}$,
        Sebastiano Bernuzzi$^6$,
        Tanja Bode$^5$,
        Bernd Br{\"u}gmann$^6$,
        Luisa~T. Buchman$^{13}$,
        Manuela Campanelli$^8$,
        Tony Chu$^{10,13}$,
        Thibault Damour$^8$,
        Jason~D. Grigsby$^6$,
        Mark Hannam$^{12}$,
        Roland Haas$^{5,13}$,
        Daniel~A. Hemberger$^{3,13}$,
        Sascha Husa$^{26}$,
        Lawrence~E. Kidder$^3$,
        Pablo Laguna$^5$,
        Lionel London$^5$,
        Geoffrey Lovelace$^{3,13,22}$,
        Carlos~O. Lousto$^8$,
        Pedro Marronetti$^{20,23}$,
        Richard~A. Matzner$^{24}$,
        Philipp M{\"o}sta$^{1,13}$,
        Abdul Mrou{\'e}$^{10}$,
        Doreen M{\"u}ller$^{6}$,
        Bruno~C. Mundim$^{1,8}$,
        Andrea Nerozzi$^{25}$,
        Vasileios Paschalidis$^4$,
        Denis Pollney$^{26,27}$,
        George Reifenberger$^{20}$,
        Luciano Rezzolla$^{1,29}$,
        Stuart~L. Shapiro$^{4,28}$,
        Deirdre Shoemaker$^{5}$,
        Andrea Taracchini$^{2}$,
        Nicholas~W. Taylor$^{13}$,
        Saul~A. Teukolsky$^3$,
        Marcus Thierfelder$^6$,
        Helvi Witek$^{17,25}$,
        Yosef Zlochower$^8$
       }

\address{$^1$ Max-Planck-Institut f{\"u}r Gravitationsphysik,
         Albert-Einstein-Institut, Am M\"uhlenberg 1,
         D-14476 Golm, Germany}

\address{$^2$ Maryland Center for Fundamental
         Physics \& Joint Space-Science Institute,
         Department of Physics,
         University of Maryland, College Park, MD 20742, USA}

\address{$^3$ Center for Radiophysics and Space Research,
         Cornell University, Ithaca, New York 14853, USA}

\address{$^4$ Department of Physics,
         University of Illinois at Urbana-Champaign, Urbana, IL 61801}

\address{$^5$ Center for Relativistic Astrophysics,
         School of Physics,
         Georgia Institute of Technology,
         Atlanta, GA 30332-0430}

\address{$^6$ Theoretisch-Physikalisches Institut,
         Friedrich-Schiller-Universit{\"a}t,
         Max-Wien-Platz 1,
         07743 Jena, Germany}

\address{$^7$ Institut des Hautes {\'E}tudes Scientifiques, 91440 Bures-sur-Yvette, France}

\address{$^8$ Center for Computational Relativity and Gravitation,
         and School of Mathematical Sciences, Rochester Institute of
         Technology, 85 Lomb Memorial Drive, Rochester, New York 14623}

\address{$^9$ Yukawa Institute for Theoretical Physics,
         Kyoto University, Kyoto, 606-8502, Japan}

\address{$^{10}$ Canadian Institute for Theoretical Astrophysics,
         University~of~Toronto, Toronto, Ontario M5S 3H8, Canada}

\address{$^{11}$ Canadian Institute for Advanced Research, 
         180 Dundas Street West, Suite 1400,
         Toronto, Ontario M5G 1Z8, Canada}

\address{$^{12}$ School of Physics and Astronomy,
         Cardiff University, Queens Building, CF24 3AA,
         Cardiff, United Kingdom}

\address{$^{13}$ Theoretical Astrophysics 350-17,
         California Institute of Technology, Pasadena, CA 91125}

\address{$^{14}$ Perimeter Institute for Theoretical Physics,
         Waterloo, ON N2L 2Y5, Canada}

\address{$^{15}$ Department of Physics, University of Guelph,
         Guelph, ON N1G 2W1, Canada}

\address{$^{16}$ Center for Computation \& Technology,
         Louisiana State University, Baton Rouge, LA 70803, USA}

\address{$^{17}$ Department of Applied Mathematics and Theoretical Physics,
         Centre for Mathematical Sciences, University of Cambridge,
         Wilberforce Road, Cambridge CB3 0WA, United Kingdom}

\address{$^{18}$ Institute of Space Sciences, CSIC-IEEC,
         08193 Bellaterra, Spain}

\address{$^{19}$ Department of Physics and Astronomy,
         The University of Mississippi, University, MS 38677, USA}

\address{$^{20}$ Department of Physics, Florida Atlantic University,
         Boca Raton, FL 33431, USA}

\address{$^{21}$ School of Mathematical Sciences
         and Complex \& Adaptive Systems Laboratory,
         University College Dublin, Belfield, Dublin 4, Ireland}

\address{$^{22}$ Gravitational Wave Physics and Astronomy Center,
         California State University Fullerton, Fullerton,
         California 92834, USA}

\address{$^{23}$ Division of Physics, National Science Foundation,
         Arlington, VA 22230, USA}

\address{$^{24}$ Center for Relativity,
         Department of Physics,
         The University of Texas at Austin,
         Austin, TX 78712}

\address{$^{25}$ CENTRA, Departamento de F\'{\i}sica,
         Instituto Superior T\'ecnico, Universidade T\'ecnica de Lisboa - UTL,
         Av.~Rovisco Pais 1, 1049 Lisboa, Portugal}

\address{$^{26}$ Departament de F\'isica,
         Universitat de les Illes Balears,
         Palma de Mallorca, E-07122 Spain}

\address{$^{27}$ Department of Mathematics,
         Rhodes University,
         Grahamstown, 6139
         South Africa}

\address{$^{28}$ Department of Astronomy and NCSA, University of Illinois at
         Urbana-Champaign, Urbana, IL 61801}

\address{$^{29}$ Institut f\"ur Theoretische Physik, 
         Max-von-Laue-Str. 1, D-60438 Frankfurt am Main, Germany}

\ead{ian.hinder@aei.mpg.de}

\begin{abstract}
The Numerical-Relativity--Analytical-Relativity (NRAR) collaboration is
a joint effort between members of the numerical relativity,
analytical relativity and gravitational-wave data analysis
communities. The goal of the NRAR collaboration is to produce
numerical-relativity simulations of compact binaries
and use them to develop accurate analytical templates for
the LIGO/Virgo Collaboration to use in detecting gravitational-wave
signals and extracting astrophysical information from them. We describe
the results of the first stage of the NRAR project, which focused on producing an
initial set of numerical waveforms from binary black holes with moderate
mass ratios and spins, as well as one non-spinning binary configuration which has a mass 
ratio of 10. All of the numerical waveforms are 
analysed in a uniform and consistent manner,
with numerical errors evaluated using an analysis code created by members of the
NRAR collaboration. We compare previously-calibrated, non-precessing 
analytical waveforms, notably the effective-one-body (EOB) and phenomenological template 
families, to the newly-produced numerical waveforms. We find that 
when the binary's total mass is $\sim 100\mbox{--}200 M_\odot$, current EOB and
phenomenological models of spinning, non-precessing binary waveforms 
have overlaps above $99\%$ (for advanced LIGO) with all of the non-precessing-binary numerical waveforms with mass ratios $\leq 4$, when maximizing over binary parameters. This implies that 
the loss of event rate due to modelling error is below $3\%$. Moreover, 
the non-spinning EOB waveforms previously calibrated to five non-spinning waveforms 
with mass ratio smaller than 6 have overlaps above $99.7\%$ with the numerical waveform 
with a mass ratio of 10, without even maximizing on the binary parameters. 
\end{abstract}



\section{Introduction}
\label{sec:intro}

A worldwide network of interferometric gravitational-wave detectors
has been operating since 2002.  This network includes
the three LIGO detectors~\cite{Abbott:2007kv}
in the United States, the French-Italian
Virgo detector~\cite{Accadia2012} in Italy, and the British-German GEO600
detector~\cite{Grote:2010zz} in Germany. After five years of collecting and analysing
data (2005-2010), the LIGO and Virgo detectors were temporarily shut
down and are currently being upgraded to the advanced LIGO and
Virgo configurations~\cite{Harry:2010zz}. These upgrades will improve
the detector sensitivities by a factor of 10. As a consequence, event
rates for coalescing binary systems will increase by a factor of one
thousand, very likely leading to the first detection and establishing
the field of gravitational-wave astronomy~\cite{Abadie:2010cf}.

The upgrades to the advanced interferometric configurations are expected
to be complete in 2015, with Advanced LIGO expected to reach full design sensitivity around 2018--2019,
although several month-long periods of observations 
are planned to take place as early as 2015~\cite{Aasi:2013wya}.
Furthermore, an underground cryogenic detector in Japan known as KAGRA is
under construction~\cite{Somiya:2011np}, and there are plans for one of the advanced
LIGO detectors to be built in India to improve sky
localization~\cite{INDIGOwebsite,Aasi:2013wya}.  During this time of
upgrades and construction, the GEO600 detector continues to operate
in the Astrowatch program to capture any potential strong events,
such as a supernova in our galaxy.  Finally, efforts to build a
gravitational-wave detector in space are under way~\cite{Amaro_Seoane:2013qna,ESALISAwebsite}.

Binary systems of compact objects ({\em compact binaries} for short), composed
of black holes and/or neutron stars, are among the most promising sources
for gravitational-wave detectors. For this class of gravitational-wave
sources, signal detection and interpretation are based on the
method of matched filtering, where the noisy detector output is
cross-correlated with a bank of theoretical templates~\cite{Dhurandhar:1992mw,
Sathyaprakash:1991mt,Balasubramanian:1995bm,Owen:1998dk,Apostolatos:1996rf,
Allen:2005fk,Finn:1992xs,Babak:2012zx,Babak:2006ty}.  A detailed
and accurate understanding of the gravitational waves radiated as the
bodies in a binary spiral towards each other is crucial not only for
the initial detection of such sources, but also for maximizing the
information that can be obtained from the gravitational-wave signals
once they are observed.

The frequency bandwidth of ground-based detectors is $\sim 10\mbox{--}10^3$ Hz, 
with best sensitivity in the $\sim
100\mbox{--}200$ Hz frequency range.  Binary neutron
stars, having masses $\sim 1 \mbox{--} 3 M_\odot$, are expected to
accumulate the majority of the signal-to-noise ratio (SNR) during the
inspiral\footnote{As a rule of thumb, an estimate of the
gravitational-wave (GW) frequency at which the inspiral ends can be obtained
from the Schwarzschild innermost-stable circular orbit, and is given by 
$f_{\rm GW} \simeq 4400/(M/M_\odot)$ Hz, $M$ being the total mass of the binary.},
with the merger at frequencies $\gtrsim 1$ kHz. In this case, the
gravitational waveform can be computed quite accurately using the
post-Newtonian (PN) approach that expands the Einstein equations in
the ratio of the characteristic velocity of the binary $v$ to the
speed of light~\cite{Blanchet2006}.
For instance, the most sensitive search for
gravitational waves from binary neutron stars with the LIGO and Virgo
detectors~\cite{Abadie:2011np} employed
non-spinning inspiral templates computed at 3.5PN order, i.e.~$(v/c)^7$~\footnote{Powers of $(v/c)^{n}$ correspond to $(n/2)$ PN
order with respect to the leading Newtonian term.}
\cite{Blanchet:1995ez,Damour:2001bu,Itoh:2003fy,Blanchet:2004ek},
which were
shown~\cite{Buonanno:2009zt,Brown:2012nn} to be sufficient for
searches of non-spinning compact object binaries of total mass up to
$12M_\odot$.

As the total mass of the binary increases, the
frequencies during late inspiral, merger, and ringdown decrease and
move into the most sensitive frequency range of the detectors.  First,
the late inspiral becomes important.  Post-Newtonian waveforms become
inaccurate in this regime where $v/c$ approaches unity, as
investigated in a series of comparisons against numerical-relativity
results~\cite{Hannam:2007ik,Boyle:2007ft,Hannam:2007wf,Boyle:2008ge,Hinder:2008kv,Campanelli:2008nk,Hannam:2010ec,MacDonald:2012mp}, and care must be taken
to develop and employ waveform templates with the correct phasing. 
As the mass increases further, the merger and eventually (at masses of
a few hundred solar masses) the ringdown
of the final black hole move into the most sensitive frequency range.
The late-inspiral and merger phases are also the most energetic parts of the
binary evolution, where up to $11\%$ of the initial total mass
of the binary is radiated in gravitational waves~\cite{Hemberger:2013hsa}.
Such high-mass
binary systems composed of black holes extend the horizon distance
of advanced LIGO and Virgo detectors from $\sim 450$ Mpc (for binary
neutron stars) to $\sim 1\mbox{--}20$ Gpc depending on the binary's
total mass, mass-ratio and spin\footnote{The horizon distance is the maximum 
distance at which advanced LIGO and Virgo can claim a detection 
for an optimally oriented binary. We compute the horizon distances 
at a single-detector SNR of 8.}.  To detect binary
black holes effectively and to take full advantage of the discovery potential of the
detectors, it is crucial to use template banks built from 
complete and accurate inspiral-merger-ringdown waveform-models.
This requires an accurate description of the non-linear, strong-field
stages of binary evolution, best provided by numerical relativity simulations.

After the dazzling breakthroughs in
2005~\cite{Pretorius:2005gq,Campanelli:2005dd,Baker:2005vv}, today several
groups are able to simulate on supercomputers the merger of compact binaries
composed of black holes and/or neutron stars (for reviews, see
e.g.~\cite{Pretorius:2007nq,Hannam:2009rd,Centrella:2010zf,Hinder:2010vn,Shibata:2011,
Sperhake:2011xk,Pfeiffer:2012pc}).
Important recent advances include simulations of black hole binaries with
precession~\cite{Tichy:2007hk,Tichy:2008du,Campanelli:2008nk,Schmidt:2010it,
O'Shaughnessy:2011fx,Mroue:2012kv,O'Shaughnessy:2012vm,
O'Shaughnessy:2012ay,Pekowsky:2012sr,Healy:2013jza}, 
large spins~\cite{Lovelace:2008tw,Lovelace:2010ne,Lovelace:2011nu,Lousto:2012es,Hemberger:2013hsa}, large
mass ratios~\cite{Lousto:2010ut,Sperhake:2011ik}, large initial 
separations~\cite{Lousto:2013oza} and large 
recoils~\cite{Campanelli:2007ew,Gonzalez:2007hi,Lousto:2011kp},
as well as particularly long and accurate simulations~\cite{MacDonald:2012mp,Buchman:2012dw,Lovelace:2011nu} and simulations in the scalar-tensor~\cite{Healy:2011ef,Barausse:2012da,Berti:2013gfa} and $f(R)$~\cite{Cao:2013} theories of gravity.
However, due to the high computational cost of numerical simulations,
template construction is currently not possible with numerical-relativity 
simulations alone.

Motivated by the construction of LIGO and Virgo
detectors, an analytical approach that combines the PN expansion and
perturbation theory, known as the {\it effective-one-body} (EOB) approach,
was introduced~\cite{Buonanno:1998gg,Buonanno:2000ef,Damour:2000we}. This
novel approach was aimed at modelling the plunge, merger
and ringdown signal of comparable-mass black holes using
physically-motivated guesses, analogies to the test-particle
limit~\cite{davis:1972,Goebel1972,Mashhoon1985} and insights from
the close-limit approximation~\cite{price_pullin94}. The EOB approach
incorporates nonperturbative and strong-field effects that are lost when
the dynamics and the waveforms are Taylor-expanded as PN series. Several
predictions of the EOB approach, notably the simplicity of the 
merger signal for non-spinning~\cite{Buonanno:2000ef} and spinning, precessing 
black holes~\cite{Buonanno:2005xu}, have been 
confirmed by the results of numerical-relativity simulations.  The 
EOB waveforms have been improved over the
years, being calibrated to progressively more accurate numerical-relativity
waveforms~\cite{Buonanno-Cook-Pretorius:2007,Buonanno:2007,Damour:2007b,Boyle:2008ge,Damour:2008te,Buonanno:2009qa,Damour:2009kr,Pan:2009wj,Pan:2011gk,
Taracchini:2012,Damour:2012ky}.

A second class of phenomenological inspiral-merger-ringdown 
waveform models has also been developed,
starting with~\cite{Pan:2007nw,Ajith:2007kx}. In this case, the
original motivation was to provide LIGO and Virgo detectors with
inspiral, merger and ringdown waveforms that could be computed
efficiently during searches and be used to observe high-mass binary
black holes. The procedure has proved sufficiently flexible
and attractive that it was also used to construct the first 
inspiral-merger-ringdown
models of non-precessing binaries calibrated to numerical-relativity
waveforms~\cite{Ajith:2011,Santamaria:2010yb}. The phenomenological
waveforms were constructed by first matching inspiral PN templates and
numerical-relativity waveforms in either the time or frequency domain,
and then fitting this hybrid waveform in the frequency domain to a
stationary phase approximation based template augmented by a
Lorentzian function for the ringdown stage. The first searches for
gravitational waves from non-spinning high-mass binary black holes with
the LIGO and Virgo detectors~\cite{Abadie:2011,Aasi:2012rja,Abadie:2012aa}
employed EOB templates calibrated to numerical-relativity waveforms to
filter the data, while phenomenological templates were used as
injection templates to study the efficiency of the search algorithm,
and have been used in LIGO-Virgo parameter estimation
studies~\cite{Aasi:2013kqa}.

All these important numerical and analytical advances have
brought us closer to the goal of observing and interpreting gravitational
waves from compact binaries. However, formidable challenges remain because
large portions of the binary parameter space are not yet covered by
accurate templates. An efficient way to span the entire parameter
space and build accurate waveforms for LIGO and Virgo searches is to
coordinate the efforts among the numerical-relativity groups and plan
simulations together with the analytical-relativity and gravitational-wave 
astrophysics communities. This is the main motivation that led to the formation of the
Numerical-Relativity--Analytical-Relativity (NRAR) collaboration in
early 2010. To this end, the U.S. National Science Foundation (NSF) made available to the
NRAR collaboration 11 million CPU hours on the Teragrid machine Kraken. The
complementary Numerical INJection Analysis (NINJA) project
was created in 2008~\cite{Aylott:2009ya,Aylott:2009tn,Ajith:2012az}.
NINJA brings together the numerical-relativity and data-analysis communities
with the goal of testing the LIGO and Virgo analysis pipelines by adding
physically realistic signals, i.e.\ the numerical-relativity waveforms,
to the detector noise in software. When pursuing these tests, analytical
template banks based on PN, EOB and phenomenological waveforms, which
are available in the LIGO and Virgo software, are used to recover the
injected signals. Those analytical template banks are the ones we aim to
improve in the NRAR collaboration. 

In this first paper, we produce an initial set of numerical waveforms from 
binary black holes with moderate mass ratios and spins, as well as one 
non-spinning binary configuration which has a mass 
ratio of 10. We provide a comprehensive technical review of current 
black-hole-binary simulation codes and methods.  We evaluate the 
numerical errors in a uniform and consistent manner, and  
then compare the numerical waveforms to previously-calibrated 
analytical waveforms to test their robustness and understand whether they need 
to be improved to better match the numerical waveforms. 

In this work, we compare only the $\ell=2,m=2$ mode of the NR waveforms and
analytical models, though other modes are
available in the NR data.  Since most of the energy is radiated in
the $\ell=2,m=2$ mode, most analytical modelling work has focused on this mode,
and it is the only mode that has been considered so far in searches
for compact binaries.  However, several studies have investigated the effects
of other modes on gravitational wave detection
algorithms~\cite{Pan:2011gk, Pekowsky:2012sr, Brown:2012nn}. Since
there can be a significant mismatch between waveforms that include
other modes and waveforms that only include the dominant mode, these
modes will likely need to be calibrated in analytical models in the
future. 

The paper is organized as follows. We start in Sec.~\ref{sec:mot} discussing the importance of 
modelling the late inspiral, merger and ringdown phases of the binary coalescence 
when searching for gravitational waves with advanced LIGO and Virgo detectors. 
In Sec.~\ref{sec:plan} we discuss
the scientific plan of the NRAR collaboration, explaining how we
selected which numerical-relativity simulations to perform from the binary parameter
space. We also review the requirements on the waveforms' length
and the accuracy requirements on the waveforms' phase and amplitude.
In Sec.~\ref{sec:codes}, we discuss the numerical codes that were used
to carry out the simulations, and provide a comprehensive review of current
methods.  In Sec.~\ref{sec:analysis} we analyse
the waveforms and compute the numerical errors.  
This is the most comprehensive error analysis to date that was applied
consistently to waveforms produced with
different numerical-relativity codes. Whereas the goal of previous 
code-comparison studies~\cite{Hannam:2009hh,Baker:2007fb} was to compare simulations of 
identical binary configurations, here we consider only one configuration simulated
by two codes, and focus instead on a uniform analysis of resolution and waveform extraction 
uncertainties across 25 waveforms produced by 9 groups using 7 numerical-relativity codes.
In Sec.~\ref{sec:NRARcomparison} we investigate how existing
analytical waveforms match the numerical waveforms produced in this
paper. Finally, Sec.~\ref{sec:concl} summarizes our main results and
gives recommendations for future projects within the NRAR collaboration.

\section{Relevance of late-inspiral, merger and ringdown phases for advanced LIGO and Virgo 
searches}
\label{sec:mot}

\begin{figure}
\centering
\includegraphics[width=0.98\textwidth, bb=10 35 750 361]{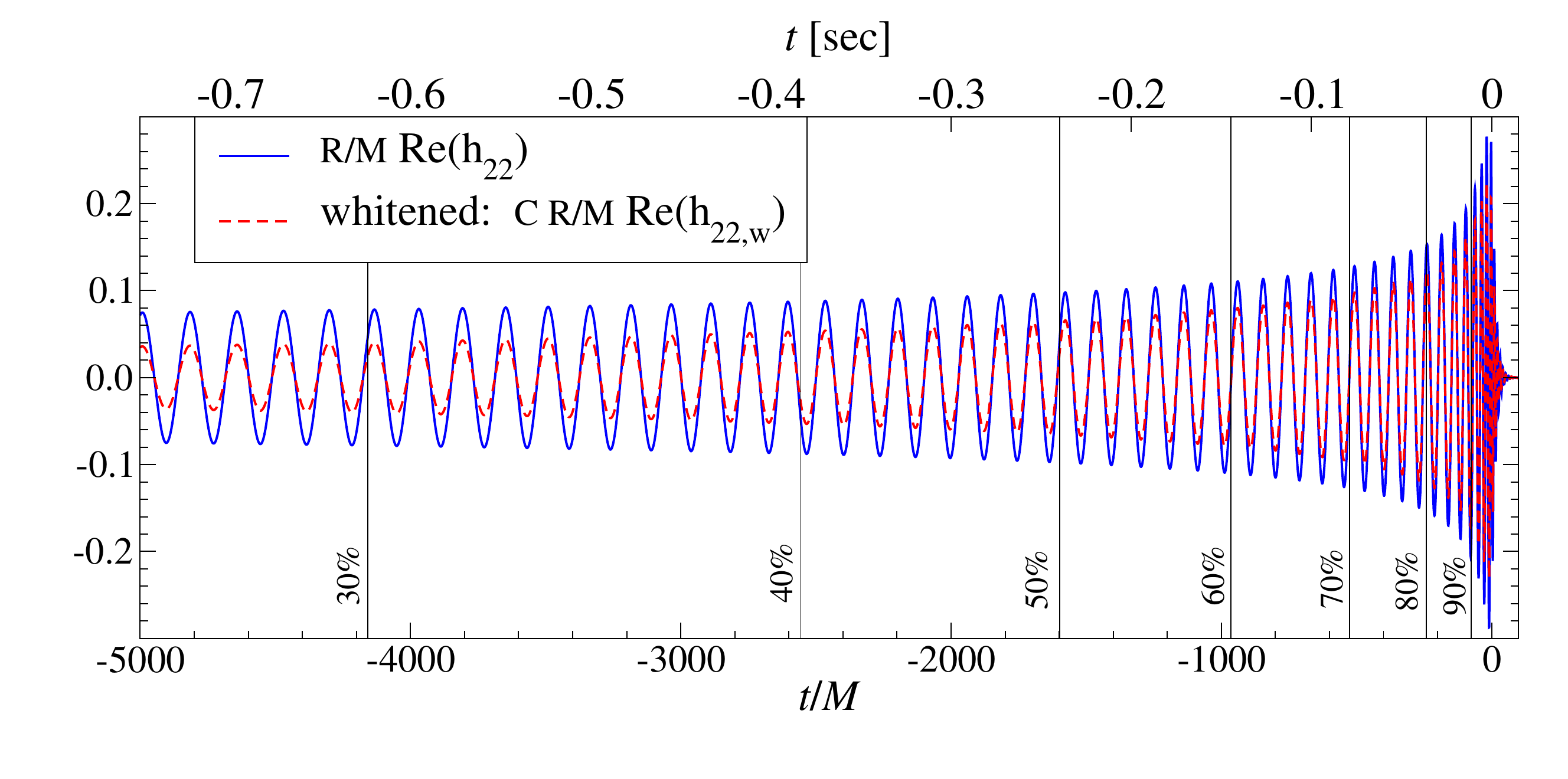}
\caption{
The $(\ell,m)=(2,2)$ mode of a binary black hole waveform with 
$M=30M_\odot$, mass ratio $q=3$ and dimensionless spins $\chi_1=-0.6$,
$\chi_2=0$.  Also shown is the same signal whitened by the square-root
of the noise spectral density of the advanced LIGO detector, and multiplied by
$C=3.1\times 10^{-24}$ for plotting on the same scale.  Shown is
 a time-domain EOB waveform model with parameters
that agree with one of the numerical waveforms presented in this paper (Case 24 (S3$-60$+00) in table~\ref{tab:Configurations}). The vertical
lines mark $10 \%$ intervals of accumulated SNR, and are labelled by the fraction of 
SNR accumulated {\em before} each line.
}
\label{fig:whit-wav}
\end{figure}

To illustrate the importance of the late inspiral and merger, 
figure~\ref{fig:whit-wav} shows a waveform for a 
binary black hole system with total mass $M = 30M_\odot$, mass ratio $q = 3$ and dimensionless
spins $\chi_1 = -0.6$,
$\chi_2 = 0$, where the minus sign indicates that the spin is oriented anti-parallel to the
system's orbital angular momentum; we shall follow this sign convention in all non-precessing
cases in the paper.  The figure also includes the waveform
{\em whitened} by the square-root of the
zero-detune high power noise spectral density of the advanced
LIGO detector~\cite{Shoemaker2009}.  The waveform parameters agree with
one of the numerical waveforms produced in the NRAR collaboration 
(Case 24 in table~\ref{tab:Configurations}).  The
vertical lines mark $10 \%$ intervals for accumulated SNR.  
For this case, the last $20$ gravitational-wave cycles
contribute $>50\%$ of the total SNR.  Therefore, for total masses 
$ M \,\gaq\, 30M_\odot$, the late inspiral, merger and ringdown waveform
is crucial for detecting the signal, and in the absence of numerical and
analytical modelling, a significant fraction of the SNR will be lost. In fact, the systematic
study in~\cite{Buonanno:2009zt} suggests that ideally inspiral-merger-ringdown waveforms
would be used in searches for binaries with $ M \,\gaq\, 12M_\odot$.

Figure~\ref{fig:whit-wav} also demonstrates that a significant
  portion of the SNR is accumulated well before merger; in this case,
  40\% of the SNR is accumulated before the last 30 cycles.  NR
  simulations often last for only $20\mbox{--}30$ gravitational-wave
  cycles, and therefore they alone will also not be sufficient to
  provide templates for LIGO and Virgo detectors.  The gravitational-wave 
  frequency at the start of the NR
  simulation provides an intuitive way to think about length
  requirements as a function of total mass. Numerical simulations can be
  rescaled to any total mass; when doing so, their dimensionless
  orbital frequency $M\Omega_{\rm in}$ at the start of the waveform is
  mapped to a gravitational-wave frequency 
\begin{equation}\label{eq:f_GW_in}
f_\text{GW, in} \sim 13\mbox{Hz}\,\frac{M \Omega_{\rm in}}{0.02}\,\left(\frac{M}{100 M_\odot}\right)^{-1}.
\end{equation}
The higher the mass $M$ of the binary, the lower this frequency.  
For a rather typical $M\Omega_{\rm in}\sim 0.02$ (20-30 gravitational-wave cycles before merger), 
(\ref{eq:f_GW_in}) indicates that the numerical waveform will cover the entire 
advanced LIGO frequency band only for $M\gtrsim 100M_\odot$.

\begin{figure}
\centering
\includegraphics[scale=0.40, bb=0 40 630 460]{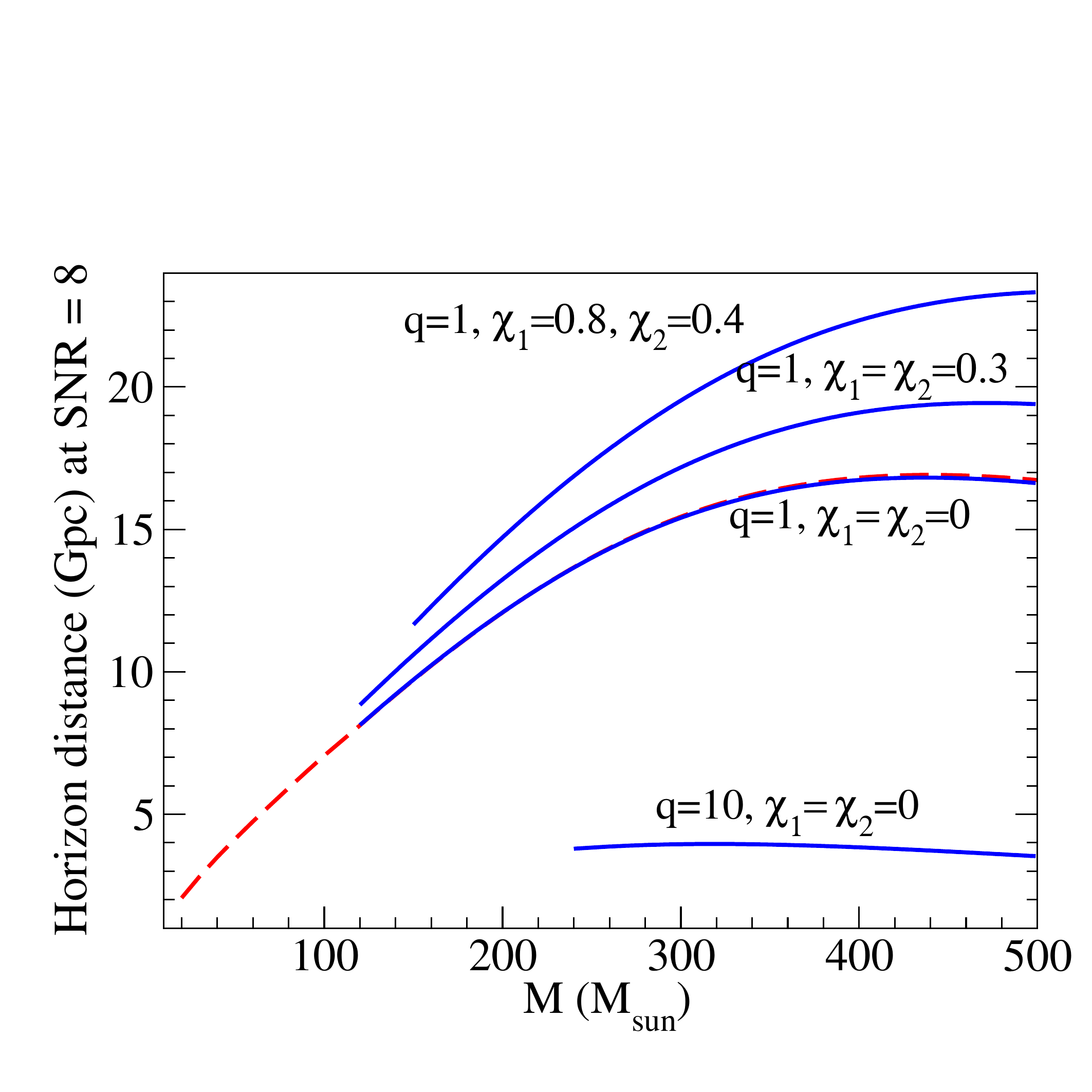}
\caption{We show the horizon distance versus the total (redshifted)
mass using some of the numerical-relativity waveforms produced in this
paper (Cases 15 (P1+80+40), 18 (S1+30+30) and 9 (R10); cf.\ table~\ref{tab:Configurations}). For mass ratio $q=1$ and no spin, we also show the numerical-relativity
waveform of~\cite{Scheel:2008rj} (solid line) and the EOB calibrated waveform
of~\cite{Pan:2011gk} (dashed line).
The numerical lines are shown only for those masses where the numerical waveform starts at frequency $10$Hz or below.
}\label{fig:HD}
\end{figure}

To better understand how numerical-relativity waveforms and the
analytical modelling are crucial to either detecting or improving the
searches for gravitational waves from binary black holes, we plot in
figure~\ref{fig:HD} the horizon distance, setting the single-detector
SNR to 8, versus the total (redshifted) mass, for a few binary mass
ratios and spins. The curves are obtained using the numerical-relativity
waveforms produced in the NRAR collaboration.  We also show the
$q=1$ non-spinning numerical-relativity waveform of~\cite{Scheel:2008rj}
and the EOB calibrated waveform of~\cite{Pan:2011gk}. In figure~\ref{fig:HD}, 
the horizon-distance curves are computed only for those total masses where numerical 
waveform starts at $\leq 10$ Hz, i.e.\ for which it covers the detector bandwidth. For the case $q=1$,
the EOB curve extends to lower masses because the
EOB waveform can easily be computed starting at lower orbital frequencies. 
Thus, unless numerical-relativity waveforms are computed starting at much
lower frequencies, the searches covering total masses $M \,\laq\, 130 M_\odot$
have to rely (in part) on analytical waveforms. The latter
have to accurately model both the merger and the last tens or hundreds of
inspiralling cycles. Longer numerical-relativity waveforms are then crucial
in testing the robustness of those analytical waveforms. 


\section{Selecting numerical-relativity simulations}
\label{sec:plan}


\subsection{Spanning the binary parameter space}

The scientific plan of the NRAR collaboration was set
up in early 2010. Considering the results that were
available at the time on analytical waveforms calibrated to
numerical-relativity waveforms in the non-spinning and spinning,
non-precessing cases~\cite{Buonanno:2007,Damour:2007b,Ajith:2007kx,
Ajith:2007xh,Boyle:2008ge,Damour:2008te,Buonanno:2009qa,Damour:2009kr,Pan:2009wj,Ajith:2011,Santamaria:2010yb}
and the limited set of numerical-relativity waveforms in the
spinning case, we proposed a general plan to span the parameter space that will
(i) underpin an initial version of the analytical model for spinning
binary systems, (ii) identify regions of parameter space where the
spinning waveforms are so sensitive to changes of parameters that they
will require further simulations, (iii) provide detailed input for the
design of those further simulations, and (iv) provide an estimate of how
many further simulations will be required to provide analytical templates 
to be used for detection (and later for parameter estimation) 
by the advanced LIGO and Virgo detectors. As we shall discuss below,  
we compare previously-calibrated analytical waveforms to the  
initial set of numerical waveforms produced by the NRAR collaboration, 
and find that current non-precessing EOB and phenomenological templates match sufficiently 
well binary systems with mild spins and mass ratios. Thus, future 
effort should focus on producing simulations of binary systems with 
larger spins and mass ratios, and stronger spin-induced precessional modulations.  

A generic gravitational waveform emitted by a binary system of black
holes with spin is described by 7 parameters: the mass ratio $q\equiv
m_1/m_2 \geq 1$ and the components of the spin vectors $\vS_1$ and $\vS_2$
at some initial time. (The spin vectors are related to the dimensionless spins $\chi_1$ and
$\chi_2$ by
$|\mathbf{S}_1| \equiv m_1^2\,\chi_1$ and
$|\mathbf{S}_2| \equiv m_2^2\,\chi_2$.~\footnote{In the non-precessing, aligned-spin cases, we
abuse notation somewhat and give $\chi_{1,2}$ a sign indicating whether the spin is
aligned (+) or anti-aligned (--) with the orbital angular momentum.} The total mass of the system, 
$M = m_1 + m_2$, provides an overall scaling, and is therefore not part of the
parameter space of binary configurations that need to be simulated. 
We also assume that the system has been circularized
by gravitational radiation during a lengthy inspiral, so its eccentricity is negligible.)
For black holes moving along
an adiabatic sequence of inspiralling, circular orbits, up to very
close to merger, these 7 parameters reduce to 6, since the initial spin
vectors $\vS_1$ and $\vS_2$ can rotate together around the initial,
orbital angular-momentum vector without changing the waveform.

At first sight, given the large number of parameters, it seems an
ambitious goal to build an accurate analytical model that covers the
entire parameter space. In fact, a naive calculation which does not take
into account any degeneracy of the parameter space and any hints from
the analytical spin modelling leads immediately to a very large number
of simulations required to span the parameter space --- for example, we
obtain $5^6\approx 15\,000$ simulations for merely five sample values in
each binary parameter's dimension.

However, exploiting degeneracies of the parameter space, a rough
estimation concludes that we need only a few hundred simulations to
achieve the goal of building an analytical model accurate enough for 
detection. One argument
goes as follows. Within the accuracy requirement for detection, in
roughly $97\%$ of the binary parameter space, the PN inspiral waveforms
are the same as one would get with a suitably-chosen single-effective-spin
black-hole binary~\footnote{The
remaining 3\% of parameter space, which
stays six-dimensional in the adiabatic approximation, has at least
one dimensionless spin larger than 0.7 and mass ratios
$q$ between about $2$ and $3$~\cite{buonanno:2004}.}
\cite{apostolatos:1994,pan:2004,buonanno:2004}.
This implies that the parameter space's dimensionality
is effectively reduced from 6 to 2 (non-precessing) or 3
(precessing)~\cite{apostolatos:1994,pan:2004,buonanno:2004}
~\footnote{We note that for some binary configurations the single effective spin can be
described by only 2 parameters~\cite{pan:2004,Ajith:2011ec}.} 
--- at least
for the inspiral waveforms. Based on experience with both the one-dimensional
non-spinning case~\cite{Ajith:2007kx,Buonanno:2007,Pan:2011gk,Damour:2012ky}
and non-precessing cases~\cite{Ajith:2011,Santamaria:2010yb,Barausse:2011kb,Taracchini:2012},
we expect that in each parameter space's dimension $\sim$5
numerical-relativity waveforms will be sufficient.  The feasibility of producing 
a non-precessing-binary model using either a handful of non-precessing waveforms 
in the small mass ratio limit~\cite{Barausse:2011kb} and only two waveforms in the equal-mass 
case~\cite{Pan:2009wj,Taracchini:2012}, or $\sim$$5^2 = 25$ waveforms in the comparable mass 
case, has been demonstrated in the EOB and phenomenological models in 
Refs.~\cite{Pan:2009wj,Ajith:2011,Santamaria:2010yb,Taracchini:2012}. This then suggests
that most of the precessing parameter space might be
covered by roughly $5^3=125$ inspiral waveforms. 
We may increase the number of simulations to compare
and validate numerical-relativity waveforms produced by different
codes. We may also include several simulations to test the degeneracy
of the parameter space predicted within the PN description of inspiral
waveforms, which went into our counting argument above, for example
as done recently in~\cite{Purrer:2013ojf}. 
Furthermore, we may need to add a certain number of merger
waveforms, because the degeneracy during the inspiral may break in the
non-linear, highly relativistic phase. However, short merger waveforms,
e.g.\ less than 10 gravitational-wave cycles, are much cheaper to simulate
numerically than lengthy inspirals. Finally, for the ringdown waveforms,
results of numerical simulations and symmetry arguments suggest that the
mass and spin of the final black hole, and hence ringdown frequencies
and quality factors, can be estimated well from the pre-merger
configuration using simple analytical formulas~\cite{Boyle:2008a,
Buonanno:2008,Rezzolla:2008,Barausse:2009,Barausse:2012}.
\begin{figure}
\centering
\includegraphics[scale=0.4]{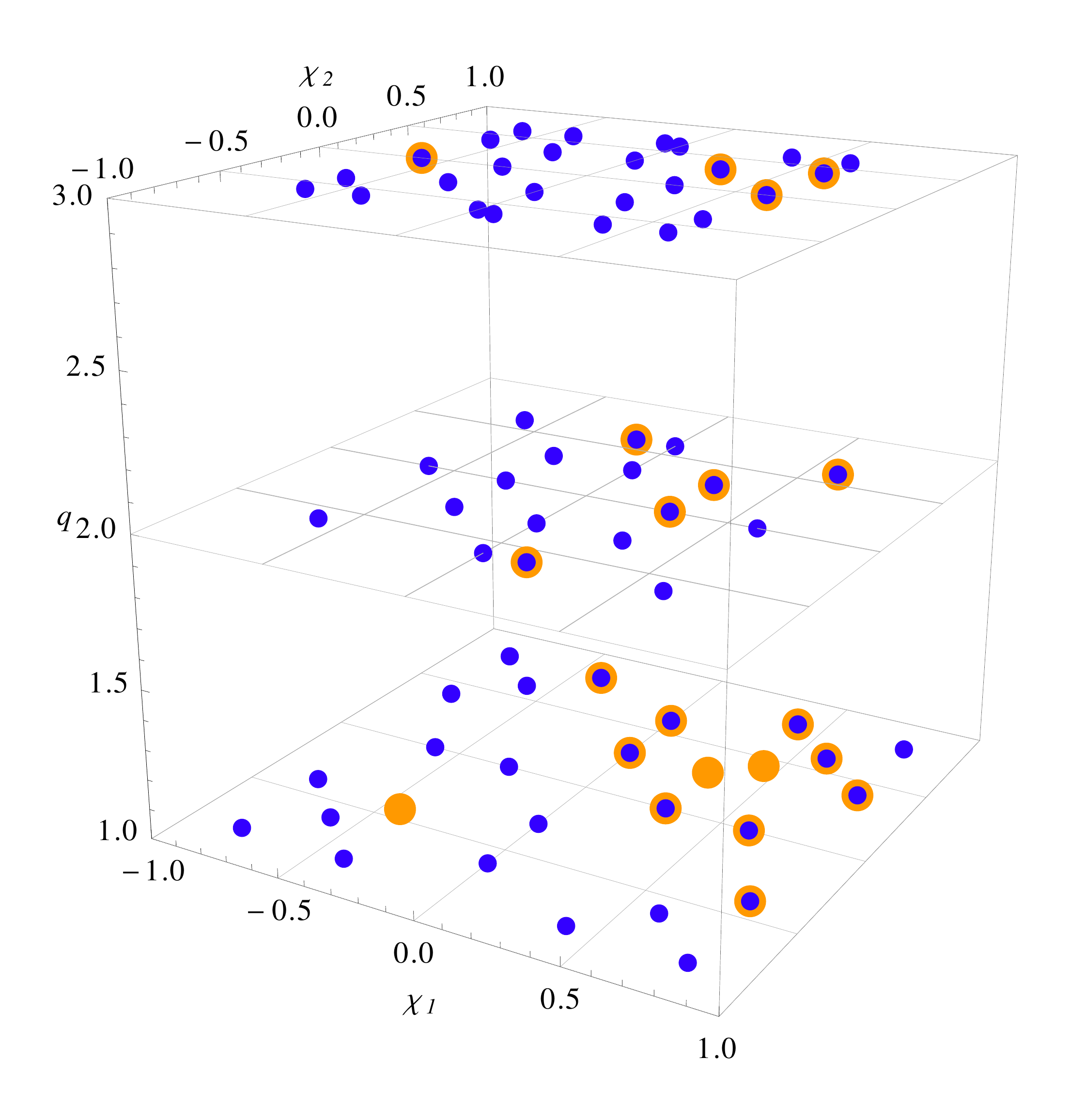}
\caption{We show in the $\chi_1$-$\chi_2$-$q$ parameter space
  the 15 spinning non-precessing configurations simulated in this paper plus
  the two equal-mass $\chi_1 = \chi_2 = \pm 0.44$ contributed simulations
  (big orange dots), as well as the remaining 43 simulations that were
  planned in the first stage of the NRAR collaboration (small blue dots).
  More details on the binary parameters of the 17 simulations that have been
  performed can be found in table~\ref{tab:Configurations}.  Note that in the $q=1$ plane,
  the points are symmetric in $\chi_1$ and $\chi_2$.} 
\label{fig:sasummary}
\end{figure}

The above considerations lead us to conclude that an initial survey
of about $200$ numerical simulations may be sufficient for detection
purposes.  However, at the time the NRAR scientific plan was established, 
the simulation of a few hundred configurations were still a demanding task
for numerical relativity. Thus, we decided to proceed in steps, dividing our initial
survey into three stages --- starting with less challenging simulations
in the first stage, extending to more generic simulations in the second
stage and concluding with several more challenging simulations in the
third stage. 

In the first stage, we planned 58 non-precessing and 19 precessing
simulations. Except for a few equal-mass binary simulations with
$\chi_1=0.8$, the binaries simulated in the first stage have mass
ratio $1\le q\le3$, component spin magnitudes $\chi_{1,2}\le 0.6$
and initial angles between the spins and the orbital angular momentum
of $0^{\rm o}, 60^{\rm o}, 120^{\rm o}, 180^{\rm o}$.~\footnote{Note that we obtain
the direction of the orbital angular momentum numerically using the Newtonian
expression applied to the coordinate positions and momenta of the black holes.
This is clearly gauge-dependent, but is the standard procedure, and we expect
it to give reasonable results when the holes are (relatively) far apart, as they are
near the beginning of the simulation, where we calculate this quantity.}
In figure~\ref{fig:sasummary} we show the 58 non-precessing configurations
in the $\chi_1$-$\chi_2$-$q$ parameter space. They cover the
parameter space quite evenly. In this paper we present 21 simulations
from the first stage (16 non-precessing and 5 precessing) and 1 from the third stage. 
Moreover, 42 non-precessing simulations from the first stage are 
under production by the SXS collaboration.
In table~\ref{tab:Configurations} we list the binary parameters of the
22 simulations produced for the NRAR project plus three
non-precessing contributed simulations. The initial data parameters used for all
 of these simulations are summarized in \ref{appendixid}.  
 Cases 12 and 25 were evolved using the same initial data in order to provide a direct
comparison between waveforms produced using different codes.  As we 
discuss below, those simulations
can be used to test and improve current analytical non-precessing models
~\cite{Santamaria:2010yb,Ajith:2011,Pan:2011gk,Taracchini:2012,Damour:2012ky},
and can be employed to start building precessing models. 

\begin{table}
\caption{\label{tab:Configurations}Configurations included in this
  study.  The waveforms are numbered in the first column, while the
  second and third columns give the simulation group and a descriptive
  label.  The label is composed of the initial of the simulation team
  (see table \ref{tab:groups}), the mass-ratio $q$, the letter `p' in the case of
  precessing binaries, and the components of the initial dimensionless
  spins along the orbital angular momentum multiplied by 100 (e.g.\ the `+15' in the label for Case 6 (G2+15--60) corresponds to
  $\hat{\mathbf{L}}\cdot\mathbf{S}_1/m_1^2 = +0.15$, where $\hat{\mathbf{L}}$ denotes the direction of
  the orbital angular momentum). We have marked  the three contributed waveforms with asterisks.
  $q$ is the mass ratio $m_1/m_2$ where $m_i$ is the mass at
  $\tjunk$.  $\mathbf{S}_i/m_i^2$ indicates the components of the dimensionless spin
  at $\tjunk$ in an orthonormal, right-handed frame where the black holes are on the $x$-axis and the orbital angular
  momentum points along the
  $z$-axis. For the non-precessing cases, where the spins are aligned with the orbital
  angular momentum, we only give the $z$-component of the spins, since the other components are
  zero. In Cases 1--4 and 23,
  the spins at $\tjunk$ were not output, so we give the initial spins. $M_f$ is the mass of
  the final black hole, and $M = m_1 + m_2$.  $|\chi_f|$ is the norm of the
  dimensionless spin of the final black hole.}
\footnotesize
\begin{tabular}{r@{$\;\;$}llrrrrr}
\br
\# & Group & Label & $q$ & $\mathbf{S}_1/m_1^2$ & $\mathbf{S}_2/m_2^2$ & $M_f/M$  & $|\chi_f|$ \\
\mr
1 & JCP & J1p$+49$$+11$ & 1 & $(-0.128, 0.171, 0.494)$ & $(0.129, -0.149, 0.106)$ & 0.941 & 0.774 \\
2 &  & J2$-15$$+60$ & 2 & $-0.150$ & $0.600$ & 0.961 & 0.611 \\
3 & FAU & F1p$+30$$-30$ & 1 & $(0.000, -0.520, 0.300)$ & $(0.520, 0.000, -0.300)$ & 0.952 & 0.704 \\
4 &  & F3$+60$$+40$ & 3 & $0.600$ & $0.400$ & 0.958 & 0.800 \\
5 & GATech & G1$+60$$+60$ & 1 & $0.603$ & $0.603$ & 0.927 & 0.858 \\
6 &  & G2$+15$$-60$ & 2 & $0.150$ & $-0.607$ & 0.962 & 0.635 \\
7 &  & G2$+30$$+00$ & 2 & $0.301$ & $0.000$ & 0.955 & 0.717 \\
8 &  & G2$+60$$+60$ & 2 & $0.601$ & $0.607$ & 0.940 & 0.839 \\
9 & RIT & R10 & 10 & $0.000$ & $0.000$ & 0.992 & 0.263 \\
10 & Lean & L4\,* & 4 & $0.000$ & $0.000$ & 0.978 & 0.472 \\
11 &  & L3$+60$$+00$ & 3 & $0.600$ & $0.000$ & 0.957 & 0.792 \\
12 & AEI & A1$+30$$+00$ & 1 & $0.300$ & $0.000$ & 0.947 & 0.732 \\
13 &  & A1$+60$$+00$ & 1 & $0.602$ & $0.000$ & 0.942 & 0.775 \\
14 & PC & P1$+80$$-40$ & 1 & $0.802$ & $-0.400$ & 0.945 & 0.744 \\
15 &  & P1$+80$$+40$ & 1 & $0.801$ & $0.400$ & 0.927 & 0.856 \\
16 & SXS & S1$+44$$+44$\,* & 1 & $0.437$ & $0.437$ & 0.936 & 0.814 \\
17 &  & S1$-44$$-44$\,* & 1 & $-0.438$ & $-0.438$ & 0.961 & 0.548 \\
18 &  & S1$+30$$+30$ & 1 & $0.300$ & $0.300$ & 0.942 & 0.775 \\
19 &  & S2$+30$$+30$ & 2 & $0.300$ & $0.300$ & 0.953 & 0.734 \\
20 &  & S3$+30$$+30$ & 3 & $0.300$ & $0.300$ & 0.965 & 0.680 \\
21 &  & S3p$+00$$-15$ & 3 & $0.000$ & $(0.260, 0.005, -0.150)$ & 0.972 & 0.536 \\
22 &  & S1p$+30$$+30$ & 1 & $(0.054, -0.514, 0.305)$ & $(0.054, -0.514, 0.305)$ & 0.937 & 0.804 \\
23 &  & S1p$-30$$-30$ & 1 & $(0.000, 0.520, -0.300)$ & $(0.000, 0.520, -0.300)$ & 0.958 & 0.638 \\
24 &  & S3$-60$$+00$ & 3 & $-0.599$ & $0.000$ & 0.978 & 0.271 \\
25 & UIUC & U1$+30$$+00$ & 1 & $0.300$ & $0.000$ & 0.947 & 0.732 \\
\br
\end{tabular}
\end{table}

\begin{table}
\caption{\label{tab:groups}Abbreviations for NR group names, as used in
  table \ref{tab:Configurations} and throughout the paper.}
\footnotesize
\centering
\begin{tabular}{ll}
\br
Abbreviation & Group name \\
\mr
JCP   & Jena-Cardiff-Palma \\
FAU  & Florida Atlantic University \\
GATech  &  Georgia Tech \\
RIT  & Rochester Institute of Technology \\
Lean  & Ulrich Sperhake \\
AEI  & Albert Einstein Institute \\
PC  & Palma-Caltech \\
SXS  & Simulating eXtreme Spacetimes  \\
& (Caltech, Cornell, CITA, CSU Fullerton)\\
UIUC  & University of Illinois, Urbana-Champaign \\
\br
\end{tabular}
\normalsize
\end{table}

The second stage will consist of a larger number of precessing binary simulations, still with
mild mass ratios and spin magnitudes. These will cover the precessing parameter
space more densely and allow the construction of
analytical precessing waveforms to be used by the advanced LIGO
and Virgo detectors to detect precessing systems. Finally, the third stage will be devoted to several
challenging simulations which either have high mass ratios ($ 3 \le q \le 15$),
large spin magnitudes ($0.6 < \chi_{1,2} < 1$), or many orbits (say, $\sim 50$
orbits before merger). It is quite important to test
the performance of analytical waveforms against numerical-relativity
waveforms that are much longer than the ones used to calibrate the
analytical waveforms. We note that the $q=10$ non-spinning simulation
completed in this paper (see table~\ref{tab:Configurations}) already
belongs to the third stage of the NRAR project. Simulations in
the third stage can also be used to test the limits of analytical models
developed in the first two stages and to guide our choices for
future simulations.

To facilitate rapid progress toward the goal of building accurate waveforms
to be used for detection by the advanced LIGO and Virgo detectors, the NSF
made available to the NRAR collaboration an allocation of 11 million CPU hours on
the Teragrid (now XSEDE) machine Kraken. This allocation, together with the computer
resources of individual numerical-relativity groups, was used to carry out
the 22 simulations presented in this paper. In addition to the waveforms produced 
specifically for this project, three previously-produced waveforms (Cases 10 (L4), 
16 (S1+44+44) and 17 (S1$-44$$-44$) in table \ref{tab:Configurations}) were included 
in the analysis.

Before concluding this section on the selection of numerical simulations, 
we notice that since the inception of the NRAR collaboration
in 2010, two algorithms have been proposed to
reduce the dimensionality of the template bank: the
singular-value-decomposition technique~\cite{Cannon:2010qh,Cannon:2011xk,Cannon:2011rj}
and the reduced-basis formalism~\cite{Field:2011mf,Caudill:2011kv,Field:2012if}. 
These algorithms could be employed in the future to span the binary parameter space 
more efficiently.


\subsection{Accuracy and length requirements}
\label{sec:acc}

Longer and/or more accurate waveforms are more costly to produce than
shorter and/or less accurate waveforms, both in terms of computational
cost and in terms of human effort. Therefore, a trade-off is
necessary between length and accuracy on the one hand, and breath of
parameter-space coverage (number of performed simulations) on the
other hand. A series of recent studies has addressed the length and accuracy 
requirements  of numerical-relativity waveforms
that are to be used for gravitational-wave data analysis.

A strict upper bound on the length and accuracy of waveforms can be
obtained if one requires that the effect of all errors that enter the
construction of gravitational waveforms do not lead to any {\em
  observable} consequences in gravitational-wave
detectors~\cite{Miller:2005qu,Lindblom:2008cm,Lindblom:2009un,Lindblom:2009ux,Lindblom:2010mh}.
This point of view was examined in~\cite{Pan:2007nw,Damour:2010zb,Hannam:2010ky,Santamaria:2010yb,
Boyle:2011dy,Ohme:2011zm,MacDonald:2011ne,MacDonald:2012mp}.\footnote{However, note that these
studies did not take the detector's calibration error into account, which can increase the waveform accuracy requirements,
as discussed in~\cite{Lindblom:2009un}.}
These studies showed that the error budget was dominated by the PN
waveforms that are needed to represent the waveforms
before the start of the numerically-computed late inspiral and merger.
In most cases the impact of the PN errors decreases with increasing order of the
PN expansion.  In the non-spinning case,
at the presently available 3.5PN order, numerical relativity 
may have to perform simulations lasting a hundred or several hundreds of
orbits (this number generally increases for larger mass ratios) in
order to completely control the errors due to the PN expansion.
Hundreds of numerical simulations of this length are impractical today.

It is worth pointing out that the criterion suggested
in~\cite{Lindblom:2009ux,Lindblom:2010mh} is a {\it sufficient}
but {\it not necessary} requirement for avoiding observable
consequences, and it does not say which of the binary parameters
will be biased and how large the bias will be.  Using the criterion
suggested in~\cite{Lindblom:2009ux,Lindblom:2010mh}, the authors
of~\cite{Pan:2011gk} concluded that the analytical template family
developed in~\cite{Pan:2011gk} would lead to systematic errors larger
than the statistical errors for SNR = 10 when $ q \,\gaq\, 6$ and the total
mass is $> 100 M_\odot$. However, a direct study~\cite{Littenberg:2012uj}
carried out with the Markov Chain Monte Carlo technique demonstrated
that the template family in~\cite{Pan:2011gk} is indistinguishable
from the numerical-relativity waveforms~\cite{Buchman:2012dw} used to
calibrate it up to SNR = 50 for the advanced LIGO detectors.

Furthermore, the very first task of gravitational-wave observatories
is the {\em detection} of signals.  Gravitational-wave searches
merely require that {\em one} of the search templates matches the {\it
exact} waveform, rather than the template with the same mass
and spin parameters as the exact waveform. This criterion of `effectualness'  
is much weaker, and~\cite{Ohme:2011zm,Hannam:2010ky} find that
approximately 10 numerical-relativity orbits are sufficient for aligned
spin binary black holes with moderate spins and moderate mass ratios;
for non-spinning binaries 10 orbits are sufficient up to $q \sim 20$.
This study also finds that in these cases parameter biases are not likely 
to affect the astrophysical information that can be inferred from observations. 
Indeed, the parameter uncertainties due to degeneracies between waveform 
parameters will in many cases be the dominant source of error for 
advanced-detector 
observations~\cite{Poisson:1995ef,Baird:2012cu,Hannam:2013uu}.

Unfortunately, none of these earlier studies is fully applicable to
our task. We would like to cover precessing systems, for which
accuracy requirements have not yet been studied.  Furthermore, instead of
simply attaching an existing PN approximant to the numerical waveforms,
we intend to {\em calibrate} analytical models to the numerical-relativity
waveforms.  Presumably, a calibration with free parameters (e.g.\ fourth
order PN coefficients) will represent the true waveform better than just
a given PN waveform.  Unfortunately, it is currently not known how much
better, because no longer numerical waveforms exist to compare against
(although such longer waveforms are becoming available, see
e.g.~\cite{MacDonald:2012mp,Mroue:2012kv}). Earlier studies that 
calibrated EOB models to numerical-relativity simulations~\cite{Boyle:2008ge,
Damour:2008te,Buonanno:2009qa,Damour:2009kr,
Pan:2009wj,Pan:2011gk,Taracchini:2012,Damour:2012ky} succeeded in
pushing the calibration errors to within the numerical truncation error
over the entire length of the numerical-relativity simulation.  Thus,
there is certainly benefit in having waveforms of comparable length
(inspiral of $\sim 30$ gravitational-wave cycles) and comparable accuracy
(phase error of $\sim 0.05$ radians during the inspiral) to these simulations
already used for calibration.  However, these
criteria are very challenging for numerical-relativity codes --- for instance, the NINJA-2
collaboration~\cite{Ajith:2012az},  had a target of $10$ usable gravitational-wave
cycles and a gravitational-wave phase accuracy of $0.5$ radians.

The discussion above shows that there are clear benefits from having higher
quality waveforms, where `higher quality' refers to longer inspirals
and smaller numerical errors.  However, attempting to increase waveform quality
too much over the status quo will be very expensive, and may be hindered
by new issues in the numerical-relativity codes which may appear when
the codes are pushed to compute waveforms of unprecedented length
and accuracy.  Therefore, we only modestly tighten the tolerances on the quality of the
numerical-relativity waveforms, sharpening them by about a factor of two relative to
what was achieved in the NINJA-2 project, noting that this corresponds to an increase of
significantly more than a factor of two in computational cost.

Specifically, we target:
\begin{itemize}
\item About 20 usable gravitational-wave cycles between $\tjunk$ and
  $\reftime$, where $\tjunk$ is the time after which the effects of
  `junk-radiation'---due to the use of non-astrophysical initial
  data---are no longer visible in the waveform, and $\reftime$ is the
  time at which the gravitational-wave frequency of the $(2,2)$ mode 
$M \omega_\mathrm{GW} = M \omref \equiv 0.2$.
\item A relative amplitude error of the $(2,2)$ mode of the gravitational
  waves of $\delta A_{22}/A_{22}\lesssim 0.01$ up to the gravitational-wave
  frequency $M\omref  = 0.2$.
\item A cumulative phase error of $\lesssim 0.25$ radians up to the
  gravitational-wave frequency $M\omref = 0.2$.
\item Orbital eccentricity $e\lesssim 0.002$.
\end{itemize}
These criteria form guidelines for the present work.  We relax some of them for
particularly challenging simulations like the mass ratio $q=10$ case.
While in general, most groups have stayed close to the guidelines to
maximize parameter-space coverage, we nevertheless have several longer
waveforms in the catalog which we will use to gain further insight into
ongoing research into length requirements for numerical waveforms.


\section{Numerical-relativity codes}
\label{sec:codes}

For the numerical solution of the Einstein field equations,
it is necessary to recast the equations in the form of an {\em initial
value} problem, where one starts from an initial snapshot of the physical system under
consideration and evolves forward in time. Approaches to achieve this
goal can be classified into (i) {\em characteristic} schemes effectively
based on the characteristics or light cones of the equations and (ii)
{\em Cauchy} or `3+1' splits where spacetime is decomposed into a
one-parameter family of spatial hypersurfaces. Simulations of black-hole
binary systems have so far only been performed with Cauchy methods and
we shall focus our discussion on these 3+1 methods. For more details on
the characteristic approach see \cite{Winicour2012}.

Quite remarkably, after nearly forty years of research,
the breakthrough in numerically evolving black-hole binaries
through inspiral and merger was achieved within a relatively
short period of time using two significantly different
3+1 frameworks: Pretorius' \cite{Pretorius:2005gq} work
employing the {\em generalized harmonic gauge} (GHG) formulation
\cite{Friedrich1985,Garfinkle:2001ni,Pretorius:2004jg} combined with 
black-hole excision, and the {\em moving punctures} technique 
developed by the Brownsville and Goddard groups \cite{Campanelli:2005dd,Baker:2005vv}
based on the Baumgarte-Shapiro-Shibata-Nakamura (BSSN) formulation
\cite{Shibata:1995we,Baumgarte:1998te}. (See \cite{Hannam:2006vv,
Hannam:2008sg,Brown:2009ki} for a
theoretical discussion of the moving puncture approach in the
Schwarzschild spacetime.)
 These methods provided the community with two independent approaches to the simulation
 of black-hole mergers, and the opportunity to validate both via a
comparison of the resulting gravitational waveforms \cite{Hannam:2009hh}.
We will briefly review the methods here and point interested readers 
to the references listed for the individual codes in table
\ref{tab:codes}.


\subsection{The Spectral Einstein Code}
\label{sec:SpEC}

The Spectral Einstein Code {\tt SpEC}~\cite{SpECwebsite} used by the SXS 
collaboration is a
pseudo-spectral multi-domain code that implements a first-order
representation~\cite{Lindblom:2005qh} of the generalized harmonic
system~\cite{Friedrich1985,Garfinkle:2001ni,Pretorius:2004jg}.
The evolution variables are the ten components of the lower index
spacetime metric $\psi_{ab}$ along with the auxiliary variables $\Pi_{ab}$
and $\Phi_{iab}$ introduced in the process of converting the original
(second differential order) system into a first-order representation.
Latin letters from the beginning of the alphabet represent space-time
indices ($a,b,c,d=0,1,2,3$), whereas latin letters from the middle of
the alphabet represent spatial indices ($i,j,\ldots=1,2,3$).
The equations are given by
\begin{eqnarray}
  \partial_t\psi_{ab}&-&(1+\gamma_1)\beta^k\partial_k\psi_{ab} 
   = - \alpha\Pi_{ab}-\gamma_1\beta^i\Phi_{iab},
   \label{eq:SpEC_psiEvol}
   \\
   \partial_t\Pi_{ab} &-& \beta^k\partial_k\Pi_{ab} 
   + \alpha g^{ki}\partial_k\Phi_{iab} - 
   \gamma_1 \gamma_2 \beta^k \partial_k \psi_{ab}
   \nonumber\\
   &=&2\alpha\psi^{cd}\bigl(  
     g^{ij} \Phi_{ica} \Phi_{jdb}
     - \Pi_{ca} \Pi_{db}
     - \psi^{ef}\Gamma_{ace}\Gamma_{bdf}
     \bigr)
     -2\alpha\nabla_{(a}H_{b)}
     \nonumber\\&&
     - {\case{1}{2}} \alpha t^c t^d \Pi_{cd}\Pi_{ab}
     -\alpha t^c \Pi_{c i} g^{ij}\Phi_{jab}\nonumber\\
     &&+\alpha\gamma_0 \bigl[2\delta^c{}_{(a}t{}_{b)}-\psi_{ab}
   t^c\bigr] ({H}_c+\Gamma_c)
   - \gamma_1 \gamma_2 \beta^i \Phi_{iab},\label{eq:SpEC_PiEvol}
   \\
   \partial_t\Phi_{iab}&-&\beta^k\partial_k\Phi_{iab}
   +\alpha\partial_i\Pi_{ab}-\alpha\gamma_2\partial_i\psi_{ab}
   \nonumber\\
   &=&{\case{1}{2}} \alpha t^c t^d \Phi_{icd}\Pi_{ab}
   +\alpha g^{jk}t^c\Phi_{ijc}\Phi_{kab}
   -\alpha\gamma_2\Phi_{iab}.\label{eq:SpEC_PhiEvol}
\end{eqnarray}
In (\ref{eq:SpEC_psiEvol})--(\ref{eq:SpEC_PhiEvol}) we used
$\alpha, \beta^i$ for the 3+1 lapse and shift and $g^{ij}$ for the
inverse of the spatial metric (which differs from the spatial
components of the inverse space-time metric $\psi^{ab}$).
Furthermore, the space-time vector $t^a$ represents the future
directed time-like unit normal to the constant$-t$ hypersurfaces,
and $\gamma_0,\gamma_1,\gamma_2$ are constraint damping parameters.
We have also made use of the four-dimensional Kronecker-delta, $\delta^a{}_b$,
the four-dimensional Christoffel symbols,
$\Gamma_{abc}$, and of their trace, $\Gamma_a = \psi^{bc} \Gamma_{abc}$.
(See~\cite{Lindblom:2005qh} for details.)  

In this formulation the
gauge source functions $ H_a = \psi_{ab} \nabla_c \nabla^c x^b$ are
freely-specifiable expressions depending on the coordinates $x^a$ and the
metric components but not on the metric derivatives.  At the beginning
of our simulations we set these so as to minimize the dynamics of the
lapse and shift.  For low-spin systems (dimensionless spin $\le0.5$)
they are transitioned smoothly in time to harmonic gauge ($H_a=0$) 
during the inspiral, while near merger we use the damped harmonic 
gauge condition
\begin{eqnarray}
  H_a &=& \mu_0\left[\ln\left(\frac{\sqrt{g}}{\alpha}\right)\right]^2
  \left[\ln\left(\frac{\sqrt{g}}{\alpha}\right)
  t_a -  \alpha^{-1} g_{ai} \beta^i \right],
\end{eqnarray}
where $\mu_0$ is a free coefficient, $g$ is the determinant of
the 3-metric and $g_{ai}$ is the spatial metric of the constant$-t$
hypersurfaces. (See \cite{Lindblom:2009tu,Szilagyi:2009qz} for
details.)  For high-spin systems (dimensionless spin $>0.5$) we transition 
to the damped harmonic gauge from the beginning of the simulation.

The {\tt SpEC} simulations presented here utilize a large number of
recent improvements, many of which were driven by the NRAR
project itself.  Two types of initial data are used: conformally flat
quasi-equilibrium initial data~\cite{Caudill:2006hw,Cook:2004kt}
for low-spin systems and superposed Kerr-Schild initial
data~\cite{Lovelace:2008tw} for higher spins.  Initial-data parameters are
tuned to achieve desired physical masses and spins with the root-finding
procedure described in~\cite{Buchman:2012dw}.  Eccentricity removal for
precessing binaries is described in~\cite{Buonanno:2010yk}.  Orbital-plane
precession is accounted for by parameterizing the rotation between
grid frame and inertial frame using quaternions~\cite{Ossokine:2013zga};
this technique works as well for non-precessing binaries as do earlier
approaches~\cite{Scheel:2006gg,Boyle:2007ft,Scheel:2008rj}, and is
therefore used for all cases.  Simulations of the inspiral phase of
conformally flat initial data use the domain decomposition described
in~\cite{Boyle:2007ft} based on constraint-damping parameters found
in~\cite{Chu:2009md}.  Simulations of the inspiral phase of superposed
Kerr-Schild initial data use a domain decomposition of touching domains
described in~\cite{Buchman:2012dw}.  Mergers and ringdowns of all inspiral
simulations, as well as the inspirals of superposed Kerr-Schild initial
data, are performed with the coordinate mappings and control systems
described in~\cite{Hemberger:2012jz}. The location of the outer boundary
is chosen as a multiple of the initial separation of the black holes and
is in the range $450$ to $650 M$.

Time stepping in {\tt SpEC} is performed with an eighth-order
Dormand-Prince time stepper~\cite{Press2007}, with adaptive time stepping
based on a fifth-order embedded updating formula.  Output at evolution
times other than the precise end of a time step utilizes the embedded
interpolation formula of the Dormand-Prince time stepper.

{\tt SpEC}'s apparent horizon finder expands the radius of the
apparent horizon as a series in spherical harmonics up to some order
$L$.  We utilize the fast flow methods developed by
Gundlach~\cite{Gundlach:1997us} to determine the expansion
coefficients.  The quasi-local spin $S$ of each black hole is computed
with the spin diagnostics described in~\cite{Lovelace:2008tw}, based
on an angular momentum surface
integral~\cite{BrownYork1993,Ashtekar-Krishnan:2004} using approximate
Killing vectors~\cite{OwenThesis,Lovelace:2008tw} of the apparent
horizons.

Gravitational waves are extracted by constructing the Newman-Penrose
scalar $\Psi_4$ on a set of coordinate spheres far from the source,
and decomposing into spin-weighted spherical harmonics of weight $-2$.
Multiple extraction spheres are used to enable extrapolation of the
waveform to infinite radius. The $\Psi_4$ extraction method used by
{\tt SpEC} is described in more detail in
Refs.~\cite{Pfeiffer:2007yz,Boyle:2007ft,Scheel:2008rj}.


\subsection{Moving Punctures Codes}
\label{sec:MPC}

The moving punctures method~\cite{Campanelli:2005dd,Baker:2005vv}
is based on a canonical `3+1' or Arnowitt-Deser-Misner (ADM)
\cite{Arnowitt:1962hi} split of the Einstein field equations
reformulated by York \cite{York1979} which recasts the equations in
terms of the spatial metric $\gamma_{ij}$ and the extrinsic curvature
$K_{ij}$ as well as a lapse function $\alpha$ and shift vector $\beta^i$,
which represent the coordinate or gauge freedom of general relativity. In
this form, the field equations appear as a set of six evolution equations
each for $\gamma_{ij}$ and $K_{ij}$,
and four constraint equations (the Hamiltonian and momentum constraints).

The ADM evolution equations are not strongly hyperbolic, and hence do not
result in a well-posed initial value problem. They therefore cannot lead
to a stable numerical discretisation using standard methods.  However, by mixing
the constraint equations into the evolution system in specific ways, the
evolution system can be made strongly hyperbolic.  The BSSN system is one
such reformulation, and has been shown~\cite{Beyer:2004sv,Gundlach:2006tw,
Witek:2010es}
to be strongly hyperbolic~\footnote{Technically, for the gauge choices
commonly used, this is true everywhere in the domain except on sets of
measure zero, but the effects of the failure of strong hyperbolicity
there are negligible.}.  In addition to the modification of the evolution
system using the constraint equations, the BSSN system also includes
a decomposition of the extrinsic curvature into trace and trace-free
parts, a conformal transformation and the introduction of the contracted
Christoffel symbols as independent variables.  These modifications make
the system suitable for moving-punctures evolutions.

The evolution variables used in the BSSN system are defined as
\begin{eqnarray}
  & \phi = \frac{1}{12} \ln \gamma, \,\,\, &
       \tilde{\gamma}_{ij} = e^{-4\phi} \gamma_{ij}, \nonumber \\
  & K = \gamma^{ij}K_{ij}, & \tilde{A}_{ij} = e^{-4\phi} \left(
       K_{ij}-\frac{1}{3} \gamma_{ij} K \right), \nonumber \\
  & \tilde{\Gamma^i} = \tilde{\gamma}^{mn}\tilde{\Gamma}^i_{mn},
  \label{eq:BSSN_vars}
\end{eqnarray}
where $\gamma$ denotes the determinant of $\gamma_{ij}$ and
$\tilde{\Gamma}^i_{mn}$ are the Christoffel symbols associated with the
conformal spatial metric $\tilde{\gamma}_{ij}$.  The evolution equations
are given by
\begin{eqnarray}
  \partial_t \tilde{\gamma}_{ij} &=& \beta^m \partial_m \tilde{\gamma}_{ij}
        + 2\tilde{\gamma}_{m(i} \partial_{j)} \beta^m - \frac{2}{3}
        \tilde{\gamma}_{ij} \partial_m \beta^m -2\alpha \tilde{A}_{ij},
        \label{eq:gamma} \\
  \partial_t \phi &=& \beta^m \partial_m \phi + \frac{1}{6} (\partial_m \beta^m
        - \alpha K),
        \label{eq:phi} \\
  \partial_t \tilde{A}_{ij} &=& \beta^m \partial_m \tilde{A}_{ij}
        + 2\tilde{A}_{m(i} \partial_{j)} \beta^m - \frac{2}{3} \tilde{A}_{ij}
        \partial_m \beta^m \nonumber \\
     && + e^{-4\phi} \left( \alpha R_{ij}
        - D_i D_j \alpha\right)^{\rm TF}
        + \alpha \left( K\,\tilde{A}_{ij}
        - 2 \tilde{A}_i{}^m \tilde{A}_{mj} \right), \\
  \partial_t K &=& \beta^m \partial_m K - D^m D_m \alpha + \alpha \left(
        \tilde{A}^{mn} \tilde{A}_{mn} + \frac{1}{3} K^2 \right),
        \label{eq:tracek} \\
  \partial_t \tilde{\Gamma}^i &=& \beta^m \partial_m \tilde{\Gamma}^i
        - \tilde{\Gamma}^m \partial_m \beta^i
        + \frac{2}{3} \tilde{\Gamma}^i \partial_m \beta^m
        + 2 \alpha \tilde{\Gamma}^i_{mn} \tilde{A}^{mn}
        + \frac{1}{3} \tilde{\gamma}^{im}\partial_m \partial_n \beta^n
          \nonumber \\
     && + \tilde{\gamma}^{mn} \partial_m \partial_n \beta^i
        - \frac{4}{3} \alpha \tilde{\gamma}^{im} \partial_m K
        + 2\tilde{A}^{im} \left( 6 \alpha \partial_m \phi
          - \partial_m \alpha \right)\,,
        \label{eq:Gamma}
\end{eqnarray}
see for example Sec.~II in Alcubierre {\em et al.}~\cite{Alcubierre:2002kk}.
Here, $D_i$ and $R_{ij}$ are the covariant derivative and the
Ricci tensor associated with the physical spatial metric $\gamma_{ij}$
and the superscript `TF' denotes the tracefree part.

When introducing a new variable (here $\tilde{\Gamma}^i$), it is necessary
to choose in which places the new variable will be used, and in which the
original will be used.  Usually, the new variable is used wherever it appears.
Either of the following two recipes yield a strongly hyperbolic
system:
\begin{itemize}
\item
Alcubierre {\em et al.}~\cite{Alcubierre:2002kk} use the variable
$\tilde{\Gamma}^i$ wherever it appears differentiated, but use the
original variable $\tilde{\gamma}^{mn}\tilde{\Gamma}^i_{mn}$ wherever
it appears undifferentiated, i.e.\ in the computation of $R_{ij}$ and
in the second and third terms of the right hand side of
(\ref{eq:Gamma}).
\item Yo {\em et al.}~\cite{Yo:2002bm} use
$\tilde{\Gamma}^i$ everywhere, but add to the right-hand side of
(\ref{eq:Gamma}) a term
\begin{equation}
  \mathcal{C}^i = -\left(\sigma + \frac{2}{3} \right)
      \left( \tilde{\Gamma}^i - \tilde{\gamma}^{mn}
      \tilde{\Gamma}^i_{mn} \right) \partial_k \beta^k\,.
\end{equation}
Here, $\sigma$ is a constant set to $\sigma=2/3$ for the simulations
performed by the GATech group and $\sigma=0$ for those
performed by the Lean group; cf.~table \ref{tab:codes}.
\end{itemize}

Note that the definition of the BSSN variables in (\ref{eq:BSSN_vars})
implies the auxiliary constraints
\begin{IEEEeqnarray}{rCl}
  \det \tilde{\gamma}_{ij} &=& 1 \,, \label{eqn:Dconstraint} \\
  {\rm tr} \tilde{A}_{ij} &=& 0\,. \label{eqn:Tconstraint}
\end{IEEEeqnarray}
The continuum evolution system in which these constraints are not
enforced is only weakly hyperbolic~\cite{Gundlach:2004jp}, leading to
instability of the finite difference scheme.  Strong hyperbolicity
results if both constraints are enforced.  Empirically (though this
has not yet been formally proven), it is sufficient to explicitly
enforce (\ref{eqn:Tconstraint}) after each time step in order to
achieve numerical stability. This is accomplished in all codes by
subtracting any residual trace contribution from $\tilde{A}_{ij}$
after each time step. In contrast, enforcing $\det
\tilde{\gamma}_{ij}=1$ appears to be optional and is implemented only
in some codes; cf.~table~\ref{tab:codes}.

A further freedom exists in the choice of variable for evolving the
conformal factor. Alternatives to the variable $\phi$
defined in (\ref{eq:BSSN_vars}) which have been suggested
in the literature are $\chi = e^{-4\phi}$~\cite{Campanelli:2005dd}
and $W=e^{-2\phi}$~\cite{Marronetti:2007wz}.

In the moving punctures approach,
the BSSN equations (\ref{eq:gamma})--(\ref{eq:Gamma})
are complemented by the `1+log' slicing
and a $\Gamma$-driver condition for the shift vector. These
are given by
\begin{eqnarray}
  \partial_t \alpha &=& \beta^m \partial_m \alpha - 2\alpha K\,, \\
  \partial_t \beta^i &=& \zeta_{\beta} \beta^m \partial_m \beta^i + \frac{3}{4} B^i\,,
             \label{eq:betat} \\
  \partial_t B^i &=& \zeta_{\beta} \beta^m \partial_m B^i + \partial_t \tilde{\Gamma}^i
             -  \zeta_{\beta} \beta^m \partial_m \tilde{\Gamma}^i -\eta B^i\,. \label{eq:Bt}
\end{eqnarray}
Here, the auxiliary variable $B^i$ is defined through (\ref{eq:betat}),
$\zeta_{\beta}$ is a constant set to $0$ or $1$ that
determines the inclusion of advection terms and
$\eta$ is a free parameter or function of dimension ${\rm length}^{-1}$.
The choices of $\zeta_{\beta}$ which yield a strongly hyperbolic system were
determined in~\cite{Gundlach:2006tw}.
Van Meter {\em et al.}~\cite{vanMeter:2006vi} suggest an alternative
first-order-in-time evolution equation for the shift vector obtained
from integration of (\ref{eq:betat}) and (\ref{eq:Bt}), viz.
\begin{eqnarray}
  \partial_t \beta^i &=& \zeta_{\beta} \beta^m \partial_m \beta^i
      + \frac{3}{4} \tilde{\Gamma}^i
      - \eta \beta^i\,. \label{eq:betat1}
\end{eqnarray}

The evolution system (\ref{eq:gamma})--(\ref{eq:Gamma})
is initialized using binary black hole Bowen-York data~\cite{Bowen:1980yu,Brandt:1997tf}
\begin{table}
\caption{Specifications of the moving-punctures codes. We list the choice
of the variable $f_{\rm conf}$ for the conformal factor, the
choice for stably evolving $\tilde{\Gamma}^i$, the
enforcement of the auxiliary constraint $\det \tilde{\gamma}_{ij}=1$,
the evolution equations for the shift $\beta^i$,
the gauge parameters
$\eta$ and $\zeta_{\beta}$, the discretization orders $n_{\rm space}$
in space
and $n_{\rm KO}$ for the Kreiss-Oliger
dissipation, the Courant factor $\Delta t / \Delta x$,
the initialization of the lapse $\alpha(t=0)$ ($\psi_\mathrm{BL}$ is the
Brill-Lindquist conformal factor given in e.g.\ (5) in~\cite{Brugmann:2008zz}),
references to the
methods employed for reduction of eccentricity in the initial data,
the range of the location of the outer boundary of the computational
domain $x_{\rm out}$,
and references containing more detailed descriptions of the
various numerical codes.
Entries of `text'
refer to a more detailed explanation given in Sec.~\ref{sec:MPC}.
\label{tab:codes}
}
\centerline{
\small
\begin{tabular}{lccccccccc}
\br
Code      &
$f_{\rm conf}$ &
$\partial_t \tilde{\Gamma}^i$ &
$\det \tilde{\gamma}_{ij}=1$ &
$\partial_t \beta^i$ &
$M\eta$ &
$\zeta_{\beta}$ &
$n_{\rm space}$ &
$n_{\rm KO}$ &
\\
\mr
JCP       &  $\chi$      & \cite{Alcubierre:2002kk}     &  yes   & (\ref{eq:betat}), (\ref{eq:Bt}) & $2.0$  & 1 & 6 & 5
\\
FAU       &  $\chi$      & \cite{Alcubierre:2002kk}     &  yes   & (\ref{eq:betat}), (\ref{eq:Bt}) & $2.0$  & 1 & 6 & 5
\\
GATech    & $\chi$       & \cite{Yo:2002bm}             & no     & (\ref{eq:betat}), (\ref{eq:Bt}) & $2.0$  & 1 & 6 & 7
\\
RIT       & $W$          & \cite{Alcubierre:2002kk}     & yes    & (\ref{eq:betat1})               & text   & 0 & 8 & 5
\\
Lean      & $\chi$       & \cite{Yo:2002bm}             & no     & (\ref{eq:betat1})               & text   & 1 & 6 & 5
\\
AEI       & $W$          & \cite{Alcubierre:2002kk}     & no     & (\ref{eq:betat}), (\ref{eq:Bt}) & $1.375$& 1 & 8 & 9
\\
PC        & $W$          & \cite{Alcubierre:2002kk}     & no     & (\ref{eq:betat}), (\ref{eq:Bt}) & $1.0$  & 1 & 8 & 9
\\
UIUC      & $\phi$       & \cite{Alcubierre:2002kk}     & yes    & (\ref{eq:betat1})               & $1.375$& 0 & 6 & 5
\\
\br
\end{tabular}
}
\vspace{0.3cm}
\centerline{
\small
\begin{tabular}{lccccc}
\br
Code      &
$\Delta t / \Delta x$ &
$\alpha(t=0)$ &
Ecc. &
$x_{\rm out}/M$ &
Refs. \\
\mr
JCP       &  0.5       & $\psi_\mathrm{BL}^{-2}$        & text                         &  $2050 - 3250$                 & \cite{Brugmann:2008zz,Husa:2007hp}
\\
FAU       &  0.25      & $\psi_\mathrm{BL}^{-2}$        & \cite{Tichy:2010qa}          & $774 - 1029$      & \cite{Brugmann:2008zz,Husa:2007hp}
\\
GATech    & 0.5        & $\psi_\mathrm{BL}^{-2}$        & \cite{Husa:2007rh}           & $410$             & \cite{Herrmann:2007ac,Healy:2008js}
\\
RIT       & 0.25       & $2/(1+1/W^2)$                  & \cite{Pfeiffer:2007yz}       & $400$             & \cite{Zlochower:2005bj,Campanelli:2005dd}
\\
Lean      & 0.5        & $\sqrt{\chi}$                  & \cite{Mroue:2010re}          & $307 - 768$       & \cite{Sperhake:2006cy,Sperhake:2007gu}
\\
AEI       & 0.45       & $\psi_\mathrm{BL}^{-1}$        & \cite{Husa:2007rh}           & $3128 - 3400$           & \cite{Pollney:2009yz}
\\
PC        & 0.45       & $\psi_\mathrm{BL}^{-1}$        & \cite{Husa:2007rh}           & $3400$                  & \cite{Pollney:2009yz}
\\
UIUC      & 0.45       & $\psi_\mathrm{BL}^{-1}$        & \cite{Husa:2007rh}           & $384$             & \cite{Etienne:2007jg,Etienne:2011ea}
\\
\br
\end{tabular}
}
\end{table}
using a spectral solver~\cite{Ansorg:2004ds} for the calculation
of the conformal factor. Particular care is required for the choice
of the Bowen-York parameter for the individual holes' linear
momentum in order to obtain an initial black-hole binary configuration
with (nearly) vanishing eccentricity. This is achieved in the individual
codes either by employing post-Newtonian or EOB model
predictions for the momenta \cite{Husa:2007rh,Walther:2009ng,
Hannam:2010ec,Tichy:2010qa} or using iterative procedures
as described in Refs.~\cite{Pfeiffer:2007yz,Mroue:2010re,
Tichy:2010qa,Buonanno:2010yk,Purrer:2012wy}.
The shift is initialized as
$\beta^i=0$ whereas the initial lapse is given as some function
of the conformal factor $\phi$.  Additionally, the bare mass parameters
in the Bowen-York initial data are determined by iterative methods to
achieve the desired physical masses, often approximated by the ADM mass
evaluated at the punctures~\cite{Brandt:1997tf,Hannam:2009ib}.

All moving punctures codes employ mesh refinement provided by {\sc
Carpet}~\cite{Schnetter:2003rb} or {\sc BAM}
\cite{Bruegmann:1997uc,Bruegmann:2003aw,Brugmann:2008zz} and use
5th order Lagrange polynomial interpolation in space and
2nd order in time (6th order and 3rd order accurate
respectively). The GATech, RIT, Lean, AEI, PC and UIUC groups use codes
based on the {\sc Cactus} framework \cite{Goodale02a,Cactuscode:web} and the Einstein Toolkit \cite{Loffler:2011ay,EinsteinToolkit:web}.
The evolution equations are evolved
using finite differencing in space
combined with
the method of lines with a fourth-order Runge-Kutta scheme for time
integration. Kreiss-Oliger dissipation~\cite{Gustafsson1995}
is added to the evolution equations, characterized by the order
$n_{\rm KO}$ which denotes the power of the spatial grid spacing
$\Delta x$ appearing in the dissipation term; see
e.g.~\cite{Zlochower:2005bj,Husa:2007hp} for details.
In addition to mesh refinement, the AEI and PC
groups also employ the {\sc Llama} multipatch
infrastructure which discretizes the wavezone by a set of six overlapping
spherical `inflated cube' grid patches \cite{Pollney:2009yz}.
In most codes, gravitational waves are
extracted by interpolating the Newman-Penrose scalar $\psifour$
onto spheres of constant coordinate radius $R_{\rm ex}$
and performing a decomposition
into multipoles using spherical harmonics of spin weight $s=-2$
(see for example Sec.~II in~\cite{Ajith:2007jx}).
The PC group extracts gravitational waves directly at future null infinity $\mathcal{J}^+$ using the method
of Cauchy-characteristic extraction (CCE) \cite{Winicour2012, Reisswig:2009us}.
Information about the black-hole properties throughout the evolution is obtained
from the apparent horizons
\cite{Thornburg:1995cp,Thornburg:2003sf} and spin estimates are obtained
through approximate Killing vectors integrated on the horizon
\cite{Campanelli:2006fy} or the relation between the horizon
area and equatorial circumference \cite{Sperhake:2009jz}.

The degrees of freedom of the individual simulations performed with the
moving punctures technique can be summarized as follows.
%
\begin{itemize}
\item The choice of evolution variable for the conformal factor.
\item The evolution of the variable $\tilde{\Gamma}^i$ using
      either the method suggested in~\cite{Alcubierre:2002kk}
      or that from~\cite{Yo:2002bm}.
\item The enforcement of $\det \tilde{\gamma}_{ij}=1$.
\item Evolution of the shift using the second-order
      equations (\ref{eq:betat}), (\ref{eq:Bt})
      or the first-order equation (\ref{eq:betat1}).
\item The choice of $\eta$ and $\zeta_{\beta}$
      in the shift condition.
\item The order of spatial finite differencing of the evolution equations and
      the order of the Kreiss-Oliger dissipation.
\item The Courant factor $\Delta t/\Delta x$, which needs to be sufficiently
      small to provide numerical stability.
\item The initialization of the lapse function.
\item The method employed for reducing eccentricity in the initial data.
\item The placement of the outer boundary of the computational domain.
\end{itemize}
In table~\ref{tab:codes}
we list the corresponding choices made in the individual
moving punctures codes. For a few choices, individual codes use
more elaborate implementations. The corresponding entries
are labeled `text' in the table and the descriptions of the
methods are given as follows.
\\[5pt]
\noindent
{\em Courant factor:} All codes use the Courant factor given in
table~\ref{tab:codes} on the inner refinement levels. However, all but
the Lean code decrease the Courant factor in the outer levels as
follows.  The RIT group decreases it on the four coarsest
(i.e.\ outermost) levels by
factors of 2, 2, 4 and 8 in outgoing order relative to the base value.
The JCP group decreases it by a factor of 2 consecutively going outwards
on the 6 outermost levels. The FAU and GATech groups
do the same on the four outermost levels, and the UIUC group on the
three outermost levels. (For these four codes, this decrease of the
Courant factor leads to a constant time step in the indicated levels.)
The AEI and PC groups, however, decrease the Courant factor by a
factor of 2 a single time on their outermost Cartesian patch and use
the resulting value on the spherical patches extending to larger
radii, as well. \\[5pt]
\noindent
{\em JCP:}
Low-eccentricity initial parameters were estimated for Case 1 (J1p$+49$$+11$)
using the method from~\cite{Husa:2007rh,Hannam:2010ec} and for Case 2
(J2$-15$$+60$) using the method from~\cite{Walther:2009ng}.
\\[5pt]
\noindent
{\em RIT:}
For the choice of the shift parameter $\eta$, the RIT group
uses a modification of the form proposed
in~\cite{Mueller:2009jx}. This modification detailed in \cite{Lousto:2010qx}
sets
$\eta(x^i,t) =  R_0\sqrt{\tilde\gamma^{ij}\partial_i W \partial_j W}
\left(1 - W^a\right)^{-b}$
with the specific choice $R_0=1.31$, $a=b=2$.
Once the conformal factor settles down to its asymptotic
form of $\psi=C/\sqrt{r} + O(1)$ near the puncture, this definition
implies that $\eta$ will have the
form  $\eta = (R_0/C^2) ( 1+ b (r/C^2)^a)$ near the puncture and
$\eta= R_0 r^{b-2} M/(a M)^b$ as $r\to \infty$. This modification is designed to
treat large mass ratio binaries.\\[5pt]
\noindent
{\em Lean:}
The Lean simulations have used a position-dependent shift parameter
$\eta$. For simulation L4, this function is given by
$M\eta_{\rm L4} = \eta_0 (r_1+r_2) / [(1+q^{-1})^{-1}r_2 + (1+q)^{-1}r_1]$,
where $r_i = |\mathbf{x} - \mathbf{x}_i|$ is the coordinate distance
of the grid point from the location of the $i$th black hole and $\eta_0=0.7$.
For simulation L3+60+00, $M\eta_{\rm L3}= [r_0^2/(r^2+r_0^2)] \times
M\eta_{\rm L4}$ where $\eta_0=1.0$ (instead of $0.7$), $r$ is the distance
of the grid point from the origin and $r_0$ a constant set to $192~M$;
cf.~\cite{Schnetter:2010cz}. Note that with this notation $\eta_0$,
unlike $\eta$, is dimensionless. 


\section{Numerical-relativity waveforms}
\label{sec:analysis}


\subsection{Strain waveforms}
\label{sec:strains}

For the purpose of binary-black-hole gravitational-wave science, it is
necessary to determine the metric strain $h$ very far from the source.
In practice, the waveform at future null infinity, $\scriplus$, is
desired.  This can be directly computed using the method of
Cauchy-characteristic extraction (CCE) \cite{Winicour2012,
  Reisswig:2009us, Reisswig:2009rx}, or obtained by computing
the waveform in the simulation at very large but finite radius, or by
extrapolating several finite-radius measurements.
Waveforms can be computed at finite radius using the Zerilli formalism
\cite{Zerilli:1970se} or from the Newman-Penrose scalar $\psifour$. A
recent investigation of the detailed relation between the two methods
can be found in \cite{Lousto:2005xu}.
The metric strain at $\scriplus$ must then be computed from these
finite-radius measurements.  While each group may use
more than one method simultaneously to compute gravitational waves, we
present results using $\psifour$ as this is the only method
implemented by all groups.

The $\psifour$ waveforms from the Palma-Caltech group are computed
directly at $\scriplus$ using CCE, while the waveforms from all the
other groups are computed at finite radii and extrapolated.

\paragraph*{Computation of strain}

Waveform modes of $\psifour$ are typically computed at several radii as
\begin{equation}
  C_{\ell m}(t,r) = \int {}^{-2}Y_{\ell m}^*(\theta,\phi) r
      \psifour(t,r,\theta,\phi) d\Omega \,,
\end{equation}
where ${}^{-2}Y_{\ell m}$ are the spherical harmonics of spin weight $s=-2$
(see \cite{Ajith:2007jx} for notation and conventions) and the star denotes
the complex conjugate.

To compute $h$ from $\psifour$ at finite radius, we use the method of
{\em fixed-frequency integration} (FFI) \cite{Reisswig:2010di}.
Unless the waveform is obtained via CCE, the waveform at $\scriplus$ is 
computed by extrapolating $h$ from several finite radii.

In the Bondi gauge\footnote{See~\cite{Lehner:2007ip} for a discussion of
the effects of waveform computation in gauges which are only approximately
Bondi.}, the strain, $h$, and $\psifour$ are related by the simple
relation $\psifour = \ddot h$.  Performing two integrations in the
time domain (even if the correct constants can be determined) can lead
to unphysical artefacts that severely contaminate the waveform, in particular
the sub-dominant modes.
Specifically, small perturbations due to numerical noise are amplified
unacceptably leading to long-term non-linear drifts in the amplitude;
see~\cite{Berti:2007fi} for examples. The
FFI method involves performing the integration in the Fourier domain,
the usual division by $\omega$ being replaced by a division by $\omega_0$
for $|\omega| < \omega_0$ to avoid the phenomenon of spectral leakage.
The method is motivated and described in detail in~\cite{Reisswig:2010di}.

The Discrete Fourier Transform (DFT) of the discretely sampled~\footnote{In
the case where the waveform is computed on a
nonuniformly-spaced grid $t_i$, the data is first interpolated onto
a uniformly-spaced grid.} $\psifour$ waveform mode, $C_j =
C_{\ell m}(t_j,r)$, $j = 0,1,\ldots,N-1$, is given by
\begin{equation}
  \tilde C_k = \frac{1}{\sqrt{N}} \sum_{j=0}^{N-1} C_j \rme^{2 \pi \rmi j k/N} \,. \label{eqn:dft}
\end{equation}
(Note that the $\ell m$ mode labels are omitted for discretized
quantities such as $C_j$ for brevity.) The strain, $H_k$, is computed
by time integration as
\begin{equation}
  \tilde H_k = \frac{\tilde C_k}{\Omega^2(\omega)} \,,
\end{equation}
where
\begin{equation*}
  \Omega(\omega) = \left\{
  \begin{array}{rl}
  \omega_0 & \text{if } |\omega| < \omega_0\,,\\
  \omega   & \text{if } |\omega| \ge \omega_0\,.
  \end{array} \right.
\end{equation*}

The choice of $\omega_0$ is guided by the principle that it should
be smaller than any important physical frequencies present in the
problem, but not so small as to introduce significant spectral
leakage~\cite{Reisswig:2009us}.  In practice, we have found acceptable
errors when using
\begin{IEEEeqnarray}{rCl}
\omega_0 &=& m\, \frac{\omega_I}{4} \,, \label{eqn:omegazero} \\
\omega_I &=& \left. \frac{d}{dt} \arg C_{22} \right|_{t=\tjunk} \,,
\end{IEEEeqnarray}
where $m$ is the spherical harmonic mode index and $\omega_I$ is the
initial frequency of $C_{22}$ measured after the junk radiation at $t
= \tjunk$.  Once a value of $\omega_0$ has been determined for $m = 2$,
values for $\omega_0$ for the other modes can be obtained from
(\ref{eqn:omegazero}).

We then compute the inverse DFT,
\begin{equation}
  H_j = \frac{1}{\sqrt{N}} \sum_{k=0}^{N-1} \tilde H_k \rme^{-2 \pi \rmi k j/N}
\end{equation}
to obtain $H_{\ell m}(t,r)$.

\paragraph*{Extrapolation of strain to $\scriplus$}

The extrapolation of finite-radius waveforms is not the
same as a direct computation at $\scriplus$, and introduces a
degree of uncertainty associated with the extrapolation error (see
Sec.~\ref{sec:erroranalysis}).  The agreement between extrapolated results
and those obtained at $\scriplus$ using CCE
was investigated in~\cite{Reisswig:2009us}.

Gravitational waves far from an isolated source propagate---to a
good approximation---along outgoing radial null geodesics.  In the
Schwarzschild spacetime, these are given by
\begin{equation}
  u = T - r_\ast(R) = \mathrm{const.} \,, \label{eqn:rettime}
\end{equation}
where $T$ and $R$ are the Schwarzschild time and areal radius coordinates,
respectively, $r_\ast$ is the Schwarzschild tortoise coordinate
\begin{equation}
  r_\ast(R) = R + 2M_\mathrm{ADM} \ln \left(\frac{R}{2M_\mathrm{ADM}}-1\right) \,,
  \label{eqn:rstar}
\end{equation}
and $M_\mathrm{ADM}$ is the mass of the Schwarzschild spacetime.
The binary black-hole spacetime and its coordinates $t, r$ approach Schwarzschild as $r \to \infty$,
so we approximate the null geodesics of the binary black-hole spacetime using
(\ref{eqn:rettime}) and (\ref{eqn:rstar}) with 
$r \approx R, t \approx T$~\footnote{All coordinate systems used by
the numerical-relativity codes are expected to satisfy the property $r \to R, t \to T$ as $r \to \infty$.
The error incurred by evaluating at a finite radius will be included in the
measure of extrapolation error (see Sec.~\ref{sec:erroranalysis}).
Better approximations here can take into account numerical measures of
the lapse and areal radius, for example, and lead to reduced errors in the
extrapolation~\cite{Boyle:2009vi}.}, and will extrapolate along
curves of constant $u$.

The waveforms are computed in the numerical codes at coordinate times which 
have no relation to the  $u=\mathrm{const.}$ curves needed for extrapolation.
We choose a set of retarded time-values $u_i$, and compute for each
radius and each $u_i$ the corresponding coordinate time $t_i(r)=u_i+r_\ast(r)$.
For each radius $r$, the finite-radius 
waveforms are interpolated to the coordinate times $t_i(r)$.
For certain gauge choices, such as the damped harmonic gauge used in 
{\tt SpEC} simulations, increased accuracy is obtained by including 
corrections to \eqref{eqn:rettime} that stem from the relatively strong variation
of the lapse and areal 
radius, especially during merger (see \cite{Boyle:2009vi} for details). 
We do not use this advanced extrapolation method in this paper to maintain 
uniformity of the analysis between simulations from different groups.

The real and imaginary parts of the $H_{\ell m}$ coefficients of a
waveform oscillate with each gravitational-wave cycle.  The complex amplitude and phase 
of these modes show less structure. These quantities are given by
\begin{IEEEeqnarray}{rCl}
  A_{\ell m}(u_i,r) &=& |H_{\ell m}(t_i(r),r)| \,, \\
  \phi_{\ell m}(u_i,r) &=& \arg H_{\ell m}(t_i(r),r) + 2 \pi n \,,
\end{IEEEeqnarray}
where $n$ is determined by continuity of $\phi_{\ell m}(u_i,r)$ in $u$.
Extrapolating quantities without oscillatory behavior (i.e.\ $A$ and $\phi$) 
reduces extrapolation errors significantly. 
For each retarded time $u_i$, and each spherical harmonic mode $(\ell,
m)$, we perform two separate linear least squares fits of the form
\begin{equation}\label{eq:fit}
  f(u_i, r) = \sum_{n=0}^{p} \frac{a_n}{r^n} \,.
\end{equation}
The left-hand-side of (\ref{eq:fit}) represents either $r$ times the
amplitude, $r A_{\ell m}(u_i,r)$, or the phase, $\phi_{\ell m}(u_i,r)$, at all extraction radii $r$ at the target retarded time
$u_i$.  The right-hand-side is the fitting polynomial with fitting
coefficients $a_n$.   

The leading order coefficient $a_0$ is then taken as the value
extrapolated to $\scriplus$ ($r \to \infty$) at order $p$. All radii
at which the whole waveform is sufficiently resolved in the numerical
simulation are included in the extrapolation.  We denote the waveform
extrapolated in this manner as $\rext_p(H_{\ell m})$.  As an
exceptional case, when $p = 0$ (no extrapolation), $a_0$ is taken to
be $f(u_i, r_\mathrm{max})$ instead of the result of the fit.

An alternative extrapolation method consists of using the results of
perturbation theory to propagate waveforms obtained at finite radius
(but in the radiation zone) to $\scriplus$. A simple explicit formula 
can be found relating $\psifour$
at $\scriplus$ with the finite radius $\psifour$ and its time integral. For
more details, see (53) in~\cite{Lousto:2010qx}. This method has
been shown to be correct for the next-to-leading  $1/r$ term in
$r\,\psifour(r,t)$ using only a single observer radius and displays a 
significantly reduced level of extrapolation 
noise~\cite{Babiuc:2010ze,Lousto:2013oza}.  
The errors produced by this method can be estimated by applying it to 
different extraction radii. We applied this method to the
$q=10$ case and found good agreement (but with significantly reduced 
noise) between the perturbative technique and the standard extrapolation
technique described above.


\subsection{Error analysis}
\label{sec:erroranalysis}

We identify and provide estimates for three distinct sources of error
in the numerical-relativity waveforms:
\begin{enumerate}
\item Finite numerical resolution;
\item Waveform measurement at finite distance from the source;
\item Computation of $h$ from $\psifour$.
\end{enumerate}

\paragraph*{Finite resolution error} results from the
conversion of the continuum Einstein equations into a discrete form
suitable for numerical solution.  The accuracy of the solution can be
increased at the expense of computational cost by decreasing the grid
spacing between the numerical grid points on which the solution is
represented (in the case of spectral codes, these are the collocation
points of the basis functions).  One can thus obtain an error estimate
by considering two such numerical solutions and the
theoretically-expected behaviour of the uncontrolled remainder terms.
However, these terms are difficult to model, since within a single
simulation, many grid-dependent approximations are used, and the
resulting errors combine in complicated ways.  For each type of code,
finite-difference and spectral, an estimate is made of the dominant
source of error and its convergence properties, and this is used to
estimate the error in the solution.

For finite-difference codes, the error is estimated using Richardson
extrapolation assuming polynomial convergence of the
solution at order $r$.  The choice of $r$ depends on the
approximations used in a given code, and is chosen by the NR group.
A quantity $q$ which converges
at order $r$ with the grid spacing $\delta$ satisfies
\begin{equation}
  q(\delta) = q_0 + C \delta^r + O(\delta^{r+1}) \,,
\end{equation}
where $q_0$ (the continuum solution) and $C$ are unknown and independent of $\delta$.  Taking
two numerical solutions with different grid spacings $\delta$, the
resulting simultaneous equations are solved for $q_0$ and $C$, and
the error in $q(\delta_2)$ due to truncation of the finite difference
approximation is
\begin{equation}
  \label{eq:FD-truncation-error}
  \sigma_\mathrm{T}(q) \equiv q(\delta_2) - q_0 = \frac{q(\delta_2)
       - q(\delta_1)}{1-(\delta_1/\delta_2)^r} + O(\delta^{r+1})\,,
       \quad\mbox{FD codes.}
\end{equation}
Note that (\ref{eq:FD-truncation-error}) assumes that the
error indeed decays like $\delta^r$, and that the higher order terms
$O(\delta^{r+1})$ are negligible.

For spectral codes in simple situations (a single domain, no time
integration, etc), the error is formally exponentially convergent in
the number of grid points.  However, due to the complexity of the
typical grids employed, and the mix of different sources of error,
this exponential convergence is usually not observed.  In tests where
the exact solution is known, the error typically behaves such that
$\sigma_\mathrm{T}(q_2) \ll \sigma_\mathrm{T}(q_1)$ if $\delta_2 <
\delta_1$ for solutions $q_1 = q(\delta_1)$ and $q_2 = q(\delta_2)$ for typical choices of grid
spacings $\delta_1$ and $\delta_2$.  Since $\sigma_\mathrm{T}(q_1) =
q_1 - q_2 + \sigma_\mathrm{T}(q_2)$, then if $\sigma_\mathrm{T}(q_2)$ can
be neglected, $\sigma_\mathrm{T}(q_1) = q_1 - q_2$ is an error estimate
on $q_1$.  Since $\sigma_\mathrm{T}(q_1) \gg \sigma_\mathrm{T}(q_2)$,
this also serves as a conservative bound on the error of $q_2$, and
is what we quote as its error:
\begin{equation}
  \sigma_\mathrm{T}(q) \equiv q(\delta_2) - q(\delta_1)\,,\qquad\mbox{\tt SpEC.}
\end{equation}

Since small dephasings can lead to large errors in $\mathrm{Re}[h]$ and
$\mathrm{Im}[h]$, given complex waveforms $h = A \exp i \phi$ computed
at different grid spacings, we quote the errors $\sigma_\mathrm{T}(A)$
and $\sigma_\mathrm{T}(\phi)$ rather than $\sigma_\mathrm{T}(h)$, as
this yields a more useful measurement.  The error is clearly identified
as a dephasing rather than a mixture of phase and amplitude errors.

\paragraph*{Finite radius error}
results from the computation of gravitational radiation at a finite
distance from the source instead of computing it at future null infinity.
All waveforms except those computed using CCE make this approximation.
To reduce this error, all waveforms are extrapolated from several finite
radii as described in Sec.~\ref{sec:strains}.  

Due to the different choices of extraction radii made by each group
for each waveform, the optimal extrapolation order $p$ varies across
the data set.  Using a lower extrapolation order leads to a larger
error, but using too high an extrapolation order leads to unwanted
{\em overfitting} of a model with a large number of degrees of freedom
in comparison to the number of data points.  We found it difficult to
construct a strict automated criteria for choosing $p$, and instead
visually inspected the behaviour of fits to $A(u,r)$ and $\phi(u,r)$
as functions of $r$ at several values of $u$.  We judged whether a
given order of extrapolation led to a reasonable fit given the number
of radii and the distance over which the extrapolation in $1/r$ was
performed relative to the difference between the minimum and maximum
radius, and whether it was possible to construct a good error
estimate.  Table \ref{tbl:extords} shows the extrapolation orders
chosen for each waveform.  Generally, second order extrapolation was
possible for $A$ for most of the waveforms, and first order for
$\phi$.  Waveforms computed at high radius $r > 200 M$ were suitable
for higher order extrapolation in $\phi$ ($p = 2$).
For three cases, the behaviour of
$A$ with radius led us to perform
no extrapolation in $A$ (denoted $p = 0$) and simply use the value from the
largest extraction radius. 

The error due to finite-radius effects is estimated by comparing the
extrapolant at different orders.  The error in the order $p$
extrapolant, $\rext_p(q)$, is estimated as
\begin{equation}
  \sigma_\mathrm{R}(q) \equiv \mathrm{min}(|\rext_p(q) - \rext_{p+1}(q)| + |\rext_{p+1}(q) - \rext_{p+2}(q)|, |\rext_{p}(q) - \rext_{0}(q)|) \,,
\end{equation}
where $q \in \{A, \phi\}$.
There are two contributions to this error estimate.
If extrapolation works well, i.e.~different extrapolation orders do not
change the results significantly, then we trust the extrapolation, and
take the variation with extrapolation order as indicative of the
error. We take the sum of the variations from the next two
extrapolation orders, to guard against coincidentally small changes,
and to ensure that extrapolation works through at least two more
orders.  If extrapolation fails (e.g.~if higher order extrapolation
fits to likely-unphysical features in the different extraction radii),
then the tentative error computed by this method will be very
large. The large value is caused by the insufficient fit, and is not a
reliable estimate of the extrapolation error. In this case, the
difference with the outermost extraction radius is used ($p = 0$).
Taking the minimum is a means to automatically switch between these
two methods.

\begin{table}
\caption{Extrapolation orders $p$ chosen for the amplitude $(A)$ and 
phase ($\phi$) for each waveform.}
\footnotesize
\centering
\begin{tabular}{llll}
\br
\#     & Label     & $p$ ($A$) & $p$ ($\phi$) \\
\mr
12--13 & A-*       & 2         & 2 \\
16--24 & S-*       &           &   \\
\mr
3      & F1p+30$-30$ & 0         & 1 \\
9      & R10              &           &   \\
11     & L3+60+00  &           &   \\
\mr
Others &           & 2         & 1 \\
\br
\end{tabular}
\label{tbl:extords}
\end{table}

\paragraph*{Strain error}
results from the computation of $h$ from $\psifour$.  The FFI method introduces
errors caused by spectral leakage and insufficiently
low choices of cutoff frequency (see~\cite{Reisswig:2010di}).  We generally
expect the cutoff frequency to be high enough that spectral leakage is
not an issue, and estimate the error due to the choice of cutoff frequency
by differentiating the strain twice to get $\psifour$, and comparing
this with the original $\psifour$.  If there was no loss of signal due
to the cutoff, we would obtain exactly the original $\psifour$.  Hence,
we use the difference as an indication of the effects of the FFI error,
and quote phase and relative amplitude errors measured in $\psifour$
here as if they were errors on the resulting $h$,

\begin{IEEEeqnarray}{rCl}
  \sigma_\mathrm{S}(\phi) &\equiv& \arg(\psifour) - \arg(\ddot h) \,, \\
  \frac{\sigma_\mathrm{S}(A)}{A} &\equiv& \frac{|\psifour|
      - |\ddot h|}{|\ddot h|} \,.
\label{eq:Errors5}
\end{IEEEeqnarray}
Better methods for estimating the error due to FFI may be developed in
the future.

\paragraph*{Non-aligned and aligned error estimates:}
In (\ref{eq:FD-truncation-error})--(\ref{eq:Errors5}), we use
differences between two waveforms to estimate the error arising from a
certain effect.  When computing these differences, we pursue two
approaches: first, we compute the differences as stated, as
functions of time $t$, where the same $t$ is used in both waveforms.
This is a non-aligned difference, and directly measures the error
in the simulation.

The second approach incorporates alignment of the two waveforms being
compared.  Because gravitational-wave detection, and matching with analytical models, always extremize over
the time-of-arrival and phase of the gravitational waveform, the
accuracy of these two degrees of freedom is less important for
those applications.  The effect of these two degrees of
freedom can be eliminated by time- and
phase-shifting one of the two waveforms being
compared relative to the other, so as to minimize a measure of their difference.

In practice, we choose an alignment window $[t_{1}, t_{2}]$, hold one waveform ($A$) 
fixed, while time- and phase-shifting the other ($B$) to minimize~\cite{Boyle:2008ge,Hannam:2010ky,MacDonald:2011ne}
\begin{equation}
  \label{eq:AlignmentMinimization}
  \Xi(\delta t, \delta \Phi) = \int_{t_{1}}^{t_{2}} \left[
    \phi_{22}^{(A)}(t)- \phi_{22}^{(B)}(t+\delta t) - 2\, \delta
    \Phi \right]^{2}\, dt \,.
\end{equation}
For a given $\delta t$, the minimization over $\delta \Phi$ can
be done analytically, reducing the problem to a one-dimensional
minimization~\cite{Boyle:2008ge}.  Waveform $B$ is then shifted in
time and phase accordingly; the phase of any $(\ell,m)$ mode is shifted
by $m\, \delta \Phi$.  Note that this transformation is appropriate for
rotations about the $z$ axis.  In significantly precessing systems,
alignments will require more general rotations~\cite{Schmidt:2010it,
BoyleOwenPfeiffer:2011,SchmidtEtAl:2012, Boyle:2013nka}.

The choice of alignment window is as much art as
science~\cite{MacDonald:2011ne, MacDonald:2012mp}, but is chosen to
be the same as the window that would be used when aligning the analytic
waveform to the numerical waveform (though such alignment is not performed
in this paper).  Thus, the resulting error measure
will describe the precision to which the analytical and numerical
waveforms can be expected to agree.  

\paragraph{Combined error estimates:}

\begin{figure}
\centerline{
\includegraphics[width=0.96\columnwidth, bb=15 50 625 465]{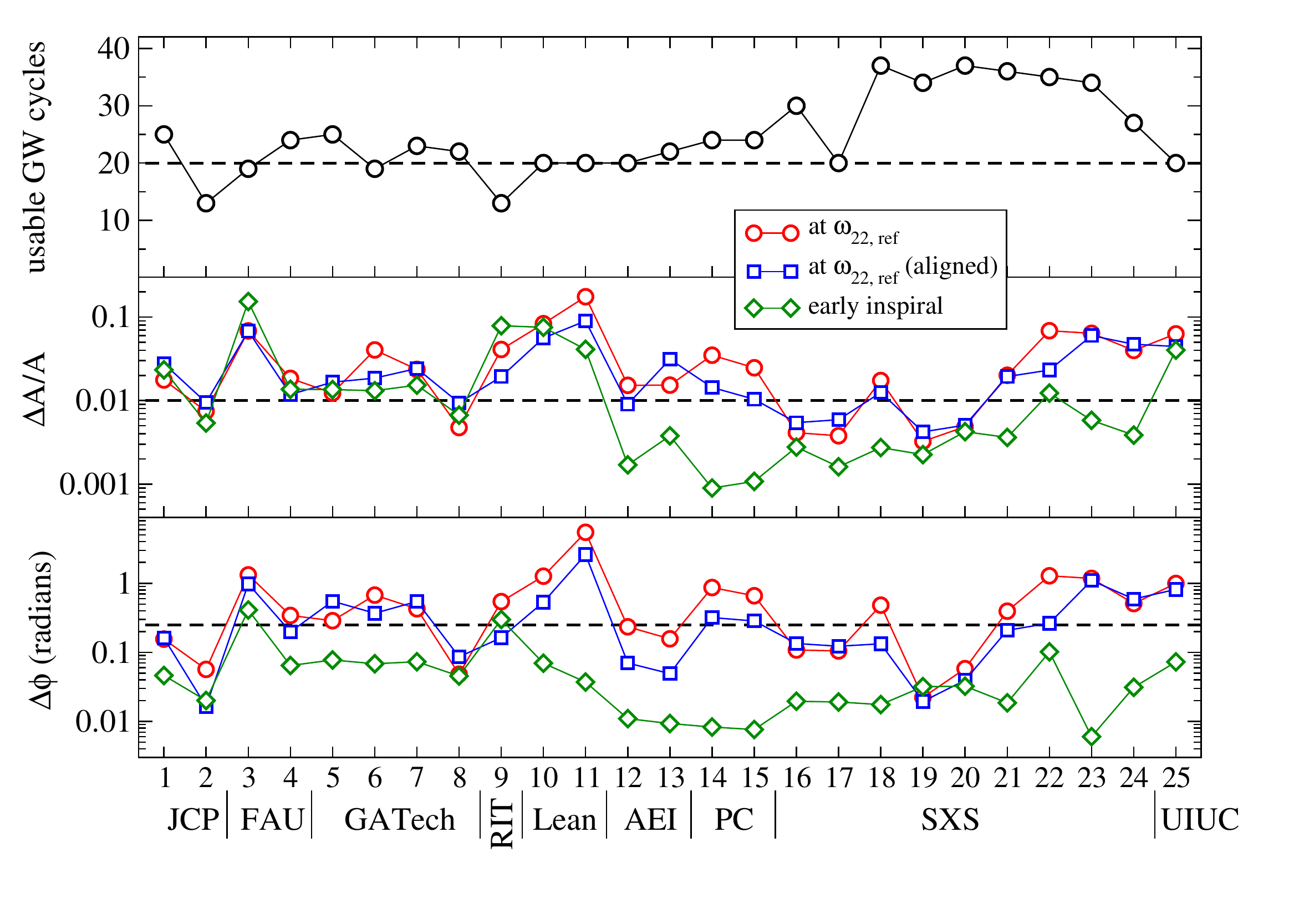}
}
\caption{\label{fig:ErrorEstimates}Error estimates for the numerical
  simulations.  The top panel shows the usable gravitational-wave cycles of each
  waveform, the middle panel the relative amplitude error of the
  $(2,2)$ mode, and the bottom panel the phase error of the $(2,2)$ mode.
  The horizontal axis indicates the case-number
     and the group name (cf.\ table~\ref{tab:Configurations}).  
     The dashed lines in each panel indicate the length and accuracy goals.}
\end{figure}

The three sources of error---truncation, finite radius and
fixed-frequency integration---each contribute to the total error in
the waveform.  Each one is computed individually without and with
alignment. We add them in quadrature to obtain our final error
estimates on the amplitude and phase of the strain waveform $h$,
\begin{equation}\label{eq:rms-error}
  \sigma(q) = \sqrt{\sigma_\mathrm{T}(q)^2 + \sigma_\mathrm{R}(q)^2
      + \sigma_\mathrm{S}(q)^2} \,,
\end{equation}
where $q \in \{A, \phi\}$.  Here, (\ref{eq:rms-error}) is
evaluated separately for the non-aligned and aligned differences, for
a total of four time-series $\sigma(q)$ per waveform.  Finally, each
of these time-series is further reduced to individual real numbers by
evaluation at both $\tjunk$ and $\reftime$ (recall that $\tjunk$ is
the time after which the effects of junk-radiation are no longer
visible in the waveform, and $\reftime$ is the time at which the
gravitational-wave frequency of the $(2,2)$ mode $M \omega_\mathrm{GW}
= M \omref \equiv 0.2$).  These
reduced error estimates are plotted in figure~\ref{fig:ErrorEstimates}
for all waveforms.  \ref{appendixerrs} contains these data in numerical
form in table~\ref{tab:ErrorEstimates}.

\paragraph{Comparison with targets:}
We now assess, with reference to figure~\ref{fig:ErrorEstimates}, the
degree to which the waveforms meet the accuracy and length targets
described in Sec.~\ref{sec:acc}.  All waveforms contain $\gtrsim 20$
usable gravitational wave cycles, apart from two, including Case 9
(R10), which requires great computational cost due to the high mass
ratio ($q = 10$).  19 out of 25 waveforms have a relative amplitude error
$\lesssim 1\%$ during the early inspiral, but all have a phase error
$\lesssim 0.25$ radians during the early inspiral.  The error at early
times is dominated by the effects of finite-radius wave extraction, as
numerical phase error in the motion of the binaries has not yet
accumulated.  All waveforms allow at least first order extrapolation
with radius in the phase, and this is shown to be sufficient to meet
the accuracy targets.  Amplitude errors during the early inspiral are
typically caused by wave extraction at too low a radius, or with too
few radii, possibly due to limitations of computational cost specific
to certain codes or binary configurations.  At the reference time
shortly before the merger, 16 out of 25 waveforms have a relative amplitude
error $\lesssim 1\%$ and 17 out of 25 waveforms have a phase error $\lesssim
0.25$ radians (we quote results for the aligned errors).  Here the
error is typically dominated by the effects of insufficient
resolution, as numerical truncation error accumulates with time,
especially for the phase.

While not all waveforms meet the accuracy targets, we chose to include
all of them in the comparison with analytical models with the
exception of Case 11, which has errors at merger significantly larger
than any of the other waveforms.  The results in
Sec.~\ref{sec:NRARcomparison} are presented with error bars computed
from these error estimates.

\paragraph{Consistency check:}
\begin{figure}
\includegraphics[scale=0.955,bb=8 7 204 127]{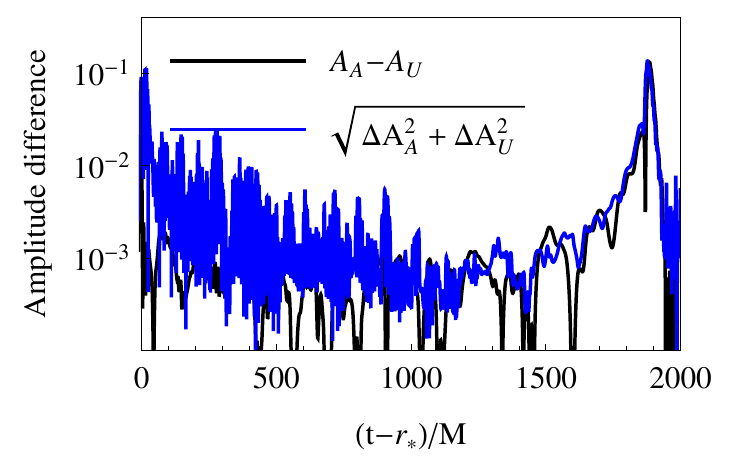}
\includegraphics[scale=0.91,bb=8 7 204 132.5]{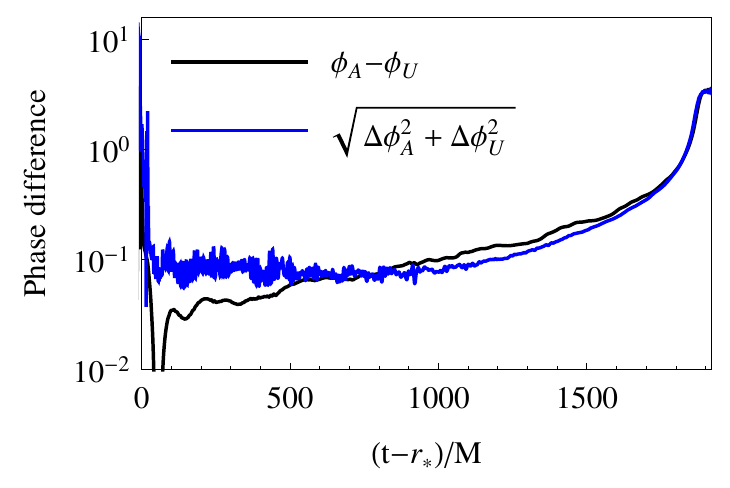}
  \caption{Comparison of the difference between the amplitudes (left)
    and phases (right) of the waveforms 12 (A1+30+00) and 25 (U1+30+00) and their
    combined error estimates. The error estimates are comparable with
    the disagreement between the waveforms, indicating that the
    results are compatible with each other. (The error estimates are not rigourous bounds, hence they do not always cover the differences completely.)
    }
  \label{fig:aucmpamp}
\end{figure}
Two of the waveforms (Cases 12 (A1+30+00) and 25 (U1+30+00)) were computed
from the same spatial initial data using two different codes. As such,
they should agree within the quoted error estimates. Figure \ref{fig:aucmpamp} shows the difference
between the amplitude and phase of each waveform compared to the
corresponding error estimate.  The error estimates are comparable with the differences,
indicating that the results are compatible.


\section{Comparison of numerical waveforms with existing analytical templates}
\label{sec:NRARcomparison}

We now compare the numerical waveforms $h_{\rm NR}$ produced in this
paper with analytical waveforms $h_{\rm AR}$ generated by existing
spinning non-precessing template models --- the time-domain SEOBNRv1
model~\cite{Taracchini:2012} and the frequency-domain phenomenological
IMRPhenomB~\cite{Ajith:2011} and IMRPhenomC~\cite{Santamaria:2010yb}
models. In the $q=10$ non-spinning case, we also compare
with analytical waveforms generated by two non-spinning EOB models ---
the EOBNRv2 model~\cite{Pan2010hz}, which has been used in the searches
of gravitational waves from coalescing binary black holes with LIGO and
Virgo~\cite{Aasi:2012rja,Abadie:2012aa}, and the recently developed IHES-EOB
model~\cite{Damour:2012ky}. The comparisons serve as both a sanity
check of the numerical waveforms and an evaluation of the analytical
models. Furthermore, they provide indications of whether the numerical
waveforms produced in this paper can be used to improve the existing
analytical models. We stress that none of the above analytical models
were calibrated using the numerical waveforms produced by the NRAR 
collaboration. However, the SEOBNRv1 model was calibrated in 
\cite{Taracchini:2012} to two 
waveforms independently generated by the SXS collaboration and later 
contributed to the NRAR collaboration, specifically Case 16 (S1+44+44) 
and Case 17 (S1$-44$$-44$) 
in table~\ref{tab:Configurations}. 

As described above, we restrict our analysis here to the
$\ell=2,m=2$ mode of the gravitational waveforms. 
We measure the difference between numerical and analytical
waveforms with the {\it unfaithfulness}~\cite{Damour:1997ub}
\begin{equation}
  \bar{\mathcal{F}} \equiv 1- \max_{t_c,\phi_c} \frac{\langle h_{\rm NR},
      h_{\rm AR}\rangle}{\sqrt{\langle h_{\rm NR}, h_{\rm NR}\rangle\langle
      h_{\rm AR}, h_{\rm AR}\rangle}}
\end{equation}
and the {\it ineffectualness}~\cite{Damour:1997ub}
\begin{equation}
  \bar{\mathcal{E}} \equiv 1
      - \max_{t_c,\phi_c,\vec{\lambda}} \frac{\langle h_{\rm NR}, h_{\rm AR}\rangle}{\sqrt{\langle h_{\rm NR},
      h_{\rm NR}\rangle\langle h_{\rm AR}, h_{\rm AR}\rangle}} \,,
\end{equation}
where we denote the time and phase of coalescence of $h_{\rm AR}$ by 
$t_c$ and $\phi_c$, and the binary parameters of $h_{\rm AR}$ by 
$\vec{\lambda}$. The dependence of $h_{\rm AR}$ on $t_c$, $\phi_c$ and $\vec{\lambda}$
has been omitted for brevity.
We define the inner product between two waveforms through 
the following integral in the frequency domain
\begin{equation}\label{matchedfilter}
  \langle h_1,h_2 \rangle \equiv 4\, {\rm Re}
      \int_0^\infty\frac{\tilde{h}_1(f)\tilde{h}_2^*(f)}{S_h(f)}df \,,
\end{equation}
where $\tilde{h}_1(f)$ and $\tilde{h}_2(f)$ are frequency-domain waveforms
\begin{equation}
\tilde{h}_k(f) = \int_{-\infty}^\infty h_k(t)\,\rme^{-2\pi \rmi f t}\,dt \quad (k=1,2)
\end{equation}
and $S_h(f)$ is the noise power spectral density of the detector.  In this
paper, we employ the zero-detuned high-power advanced LIGO noise curve
\texttt{ZERO\_DET\_HIGH\_P}~\cite{Shoemaker2009}. The
unfaithfulness of analytical waveforms is
the normalized inner product minimized over $t_c$ and $\phi_c$ and
it is related to the bias in measuring the binary parameters. The
ineffectualness is also minimized over
$\vec{\lambda}$ and quantifies the efficiency in detecting
gravitational-wave signals. Although here we are mainly interested 
in understanding whether the existing template families are effectual in
detecting the numerical waveforms given in table~\ref{tab:Configurations}, we also 
want to study the consistency of these models against the numerical 
waveforms over the entire frequency range. This will tell us what region future 
improvements should focus on. Thus, both unfaithfulness and ineffectualness 
give us important, complementary information. Because it is computationally expensive to
minimize over $\vec{\lambda}$ when using the time-domain EOB models, 
all the plots in this section will show the unfaithfulness, 
and for only the few cases for which the unfaithfulness is 
larger than $1\%$, we calculate the ineffectualness. 

The definition of $\langle h_1,h_2\rangle$ given in
\eqref{matchedfilter} involves an integral over frequency from
$0$ to $\infty$. In reality, since the noise power spectral density of the
detector has a sharp low-frequency cutoff due to seismic noise, it is
safe to start the integral at this cutoff frequency, which is $10$~Hz
for the \texttt{ZERO\_DET\_HIGH\_P} noise curve of the advanced LIGO
detector. However, for a reasonable range of the total mass $M$ of a
binary system, which we choose to be $20M_\odot\mbox{--}200M_\odot$
in this section, the numerical waveforms do not always have a small enough
initial dimensionless frequency to start at a physical frequency of $10$~Hz. In fact,
the numerical waveform with the largest initial (dimensionless) frequency (Case 9 (R10)) 
starts at $(220M_\odot/M)\times 10$ Hz, while the one with the smallest initial
frequency (Case 23 (S1p$-30$$-30$)) starts at $(90M_\odot/M)\times 10$ Hz (see also 
(\ref{eq:f_GW_in}) above and discussion around it). Here we do not build hybrid 
waveforms by joining together numerical-relativity waveforms to 
PN-approximants~\cite{Ajith:2012az}, 
because we are interested in assessing the closeness of the analytical
waveforms to numerical waveforms and do not want to spoil the result by
introducing errors due to the disagreement between the analytical
waveforms and the PN-approximants used to build the hybrid waveforms.

\begin{figure}
  \centering
  \includegraphics[width=0.7\textwidth,bb=30 47 590 395]{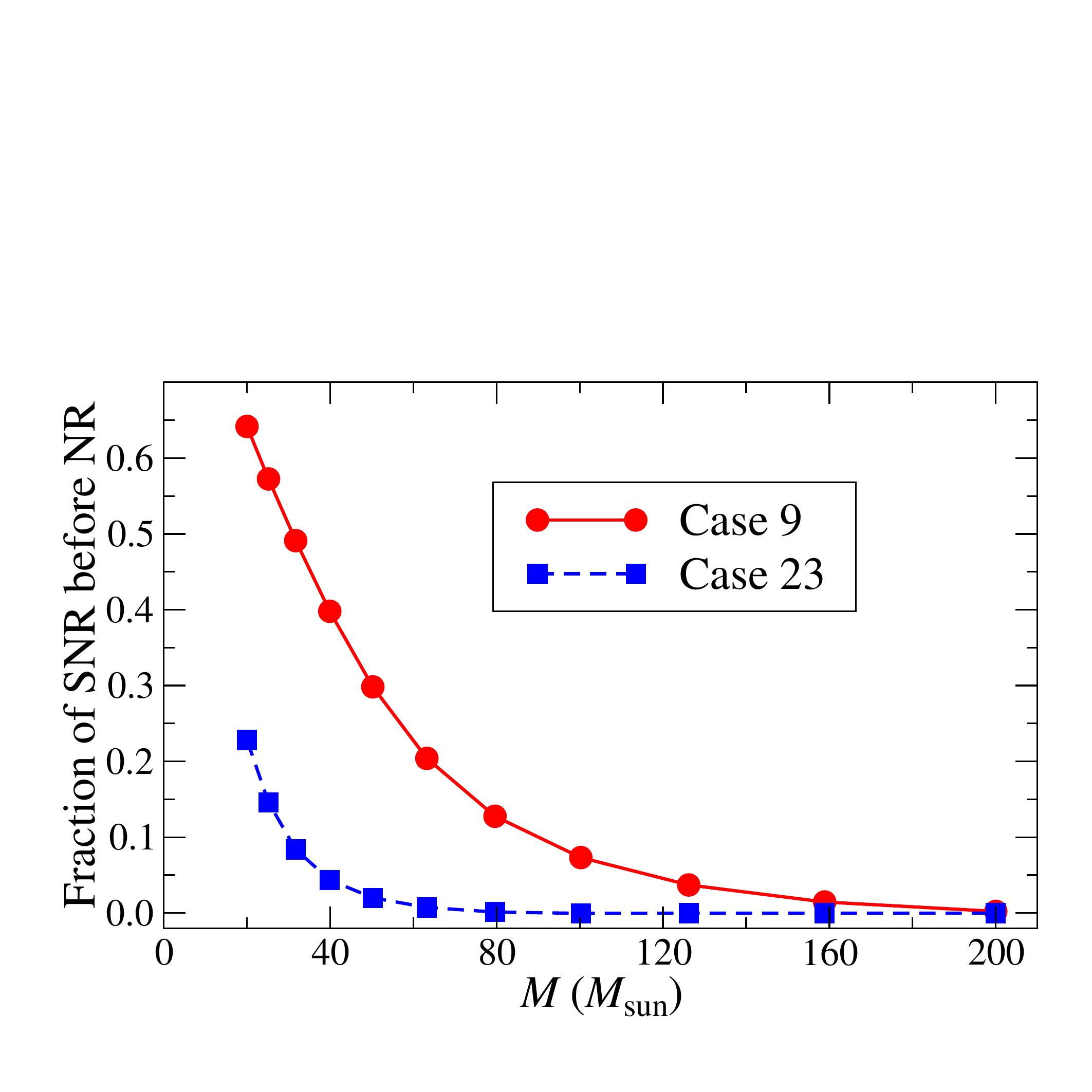}
  \caption{Fraction of SNR accumulated \emph{before} the numerical waveform starts.  
    For NR-only overlap calculations, we begin the
    overlap integral given in (\ref{matchedfilter}) at the frequency where
    the numerical waveform begins, ignoring the fraction of SNR shown in this figure, which would be
    accumulated at lower frequencies. The losses shown in this figure are calculated using SEOBNRv1
    waveforms for binary configurations 9 (R10) and 23 (S1p$-30$$-30$)  in
    table~\ref{tab:Configurations}. The low-frequency truncations are
    at $(220M_\odot/M)\times 10$ Hz and $(90M_\odot/M)\times 10$ Hz
    for Cases 9 and 23, respectively.}
  \label{fig:SNRloss}
\end{figure}
To reduce artefacts when the numerical waveforms start in the detector
bandwidth, we truncate the analytical waveforms in the
time domain when they reach the starting frequency of the numerical waveform and taper all
waveforms using the Planck-taper window
function~\cite{McKechan:2010kp}.
The width of the window function is set to the length of the numerical waveforms, starting 
from $t = t_0$, after which the effects of junk radiation are no longer important. 
The window function smoothly rises from $0$ to $1$ in the first $200M$ and falls
from 1 to 0 in the last $20M$ of the waveforms. The unphysical junk radiation is therefore completely excluded from the comparison. 
For the integral in \eqref{matchedfilter}, we choose as our low-frequency cutoff
 $f_{\rm low}$ the frequency at which the numerical 
  waveforms start.  This low-frequency cutoff is chosen separately
  for each numerical waveform, and can be as high as 110Hz.  Since these
  low-frequency cutoffs are substantially inside the sensitivity
  range of the \texttt{ZERO\_DET\_HIGH\_P} noise curve used here, we must be
  careful when interpreting the unfaithfulness and ineffectualness
  results.

  In the absence of longer numerical waveforms, we cannot exactly quantify the 
  effect of choosing the starting frequency of the numerical waveform as our
  low-frequency cutoff. However, we can at least investigate and estimate 
the effect of neglecting the SNR below the starting frequency of the 
numerical waveform. In figure~\ref{fig:SNRloss} we show the fraction of SNR
  lost due to the low-frequency cutoff for the two numerical waveforms
  with the lowest and highest starting
  frequencies. Figure~\ref{fig:SNRloss} is computed using SEOBNRv1
  waveforms having the same physical parameters as the numerical
  waveforms.  We can see that for the waveform with the lowest
  initial frequency (Case 23 (S1p$-30$$-30$)), the loss of SNR is less than 
  a quarter when $M$ is above $20M_\odot$ and less than $1\%$ when $M$ is 
above $100M_\odot$, indicating that the 
  amount of signal lost by truncating the waveforms is 
  relatively small. We therefore expect
  the results calculated from the low-frequency truncated integrals to
  be reliable estimates of the exact unfaithfulness and
  ineffectualness, at least when $M$ is above $100M_\odot$. 
  For the waveform with the highest initial
  frequency (Case 9 (R10)), however, the loss of SNR can be more than half 
  when $M$ is around $20M_\odot$. Thus, in this case, the unfaithfulness and 
  ineffectualness results calculated from the low-frequency truncated integrals 
  do not allow us to draw definitive conclusions about whether the analytical 
  waveforms are sufficiently accurate for detection and/or measurement 
  purposes for low-mass binaries. 
Nevertheless, we believe that even when the loss of SNR is high, 
say $\gaq\,10\%$,  
as in Case 9 (R10) for $M \leq 100 M_\odot$, the results 
obtained with the low-frequency truncated integrals are still
meaningful for understanding and improving the modeling errors in the frequency band
covered by the current numerical waveforms. 

The reason is the following. When the total mass is $100M_\odot$($200M_\odot$), 
the numerical waveform for Case 9 (R10) covers a frequency range of 
$22$--$135$Hz ($11$--$68$Hz) and the detector noise power spectral density 
$S_h(f)$ decreases by more than one (two) order(s) of magnitude 
going from the low end to the high end of the frequency range. 
Thus, when $M \geq 100M_\odot$, the unfaithfulness and ineffectualness results are 
particularly sensitive to the differences between the numerical and analytical waveforms at high 
frequency during the merger-ringdown stage. On the contrary, when the total 
mass is $20M_\odot$, $S_h(f)$ changes by less than a factor of $1.5$ in the frequency range of
  $110$--$680$Hz and the unfaithfulness and ineffectualness results provide us with 
a broader measure of the overall agreement of
  the numerical and analytical waveforms during inspiral, merger 
and ringdown. Thus, when the total mass decreases from 
  $100 M_\odot$ to $20M_\odot$, the unfaithfulness and ineffectualness are 
  increasingly more sensitive to modeling errors at lower frequency. If  
  the unfaithfulness and ineffectualness are high, it means that the analytical
  modeling of the last stages of inspiral (i.e.\ $20\mbox{--}30$ gravitational-wave cycles before 
  merger) is not very reliable. In summary, unfaithfulness and ineffectualness results 
 computed with the low-frequency truncated integral may not be used to draw definitive 
conclusions for detection and parameter-estimation purposes when $M \leq 100M_\odot$,  
but they do provide us with important information on the modeling errors of the 
analytical waveforms during the last stages of inspiral, and can guide us 
in improving the models in the future. 

Finally, we have carried out the following test to further check the use of the 
low-frequency truncated integral. We have applied the procedure proposed 
in~\cite{Ohme:2011zm} and have extended both the numerical and analytical waveforms (SEOBNRv1 model) at low frequency attaching to them the {\it same waveform}, so that they span the entire 
frequency band of advanced LIGO. Using those extended waveforms, we have computed 
the unfaithfulness and compared it with the unfaithfulness derived 
with the low-frequency truncated integral. We have found very small differences between these 
two methods that have no impact on any conclusion of this paper. 
Therefore, we henceforth restrict our discussion to the unfaithfulness and ineffectualness  
computed with the low-frequency truncated integral for total masses $20\mbox{--}200 M_\odot$ .

\begin{figure}
\includegraphics[width=\textwidth, bb=20 20 700 340]{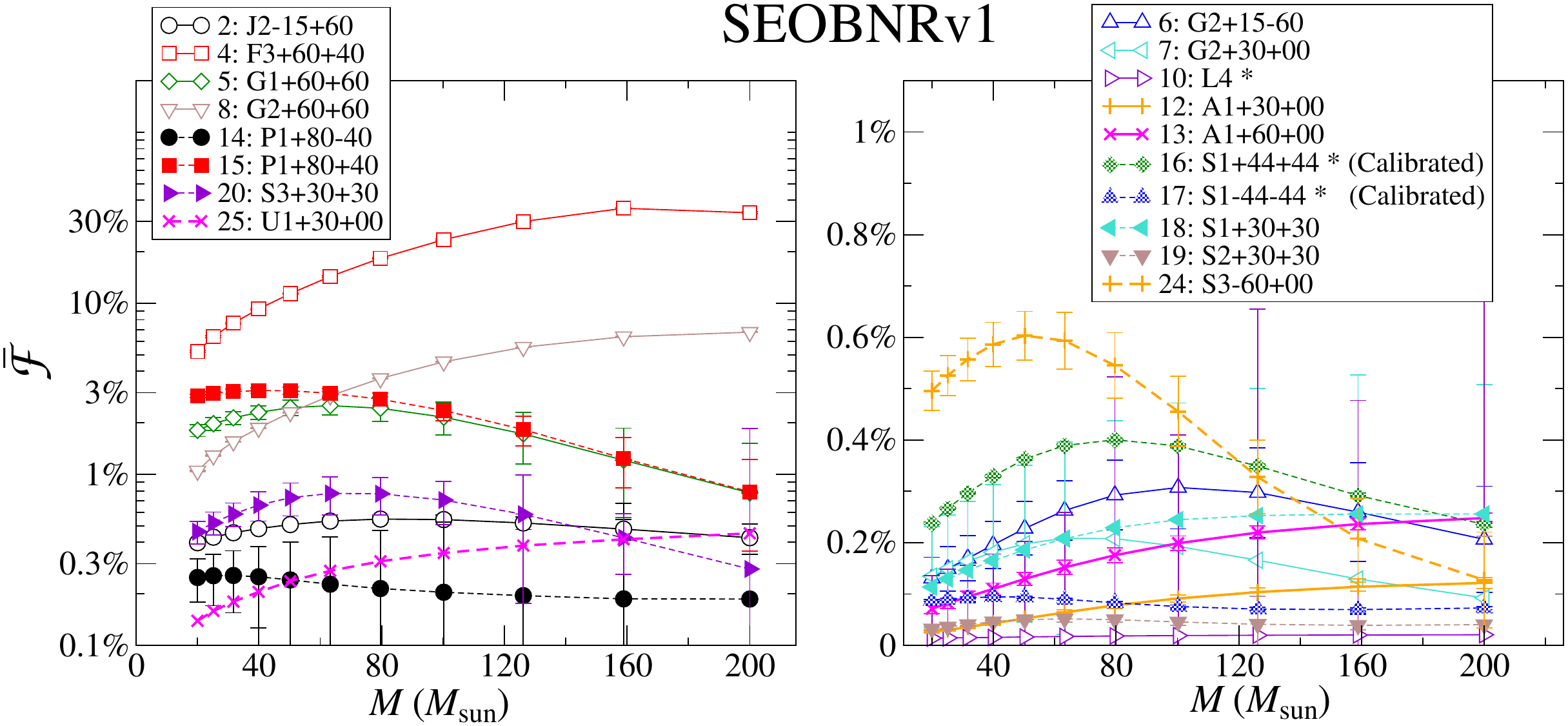}
\caption{Unfaithfulness $\bar{\cal F}$ of the
    SEOBNRv1 waveform model compared to the numerical simulations.  The
    horizontal axis shows the total mass, the vertical axis $\bar{\cal F}$ in percent.
The waveforms are split between the two panels for readability.  
Note that for $M \lesssim 100 M_\odot$ the plot 
{\it disregards} the low-frequency portion of the waveform which is 
below the starting frequency of the numerical waveform yet of substantial 
contribution to the SNR (cf. Sec.~\ref{sec:NRARcomparison} and 
figure~\ref{fig:SNRloss}). Numerical relativity error-estimates 
$\delta{\bar{\cal F}}$ on the computed unfaithfulness values are 
plotted, except for Cases 18 and 25 for which $\delta{\bar{\cal F}} > 20\%$. 
Note that the SEOBNRv1 model was calibrated to the contributed Cases 16 (S1+44+44) and 
17 (S1$-44$$-44$) shown in the right panel with shaded symbols.}
\label{fig:mm}
\end{figure}
%


\subsection{Results using the effective-one-body waveform models}
\label{sec:EOB}

\begin{figure}
\vspace{1.2cm}
\centering
\includegraphics[scale=0.35,bb=18 10 600 445]{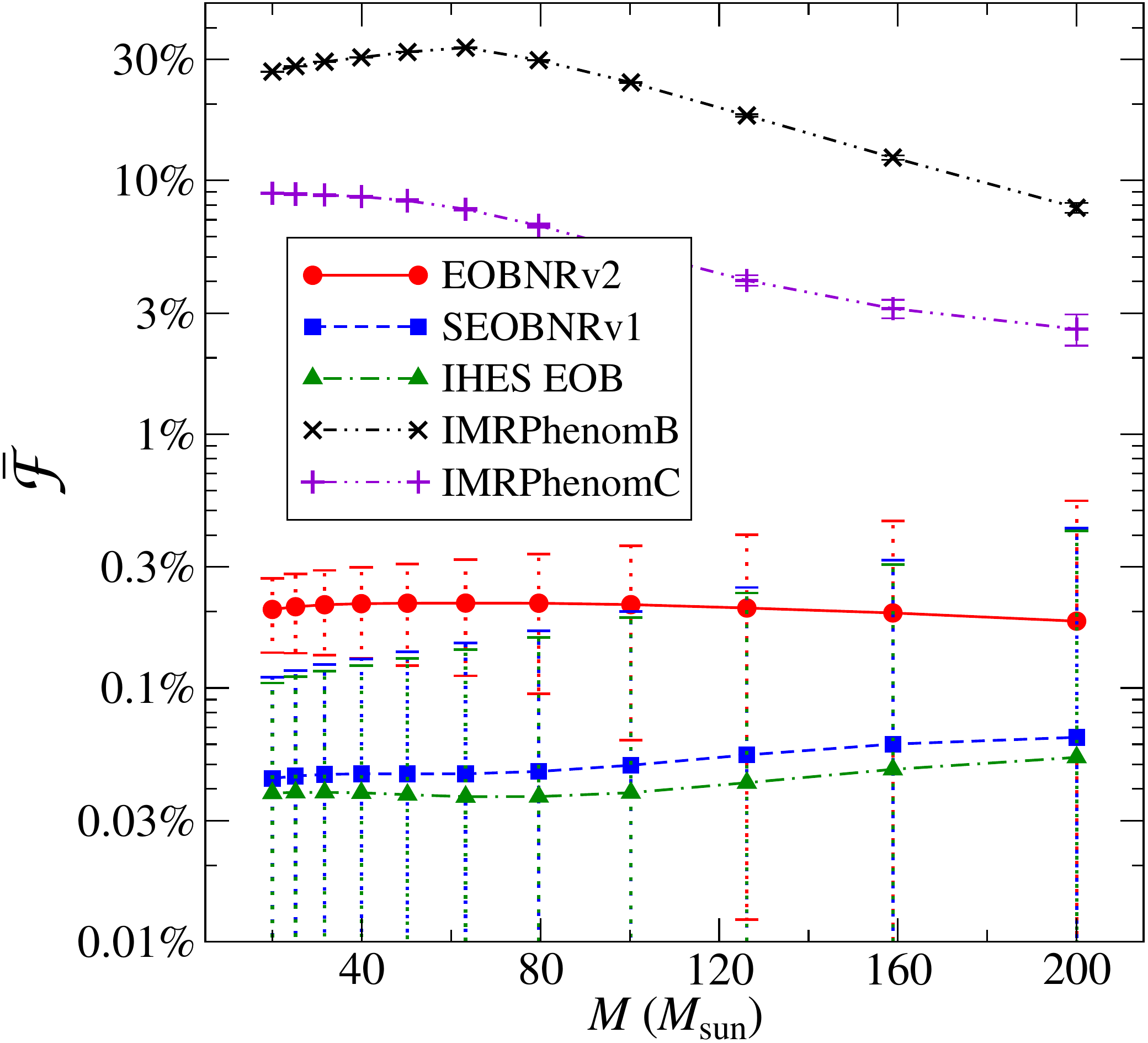}
\caption{Unfaithfulness of analytical non-spinning $q=10$ waveforms generated by 
IMRPhenomB, IMRPhenomC and three versions of EOB models with the numerical non-spinning $q=10$ waveform (Case 9 (R10)).
  The numerical error bars are the same for all sets of
  results.
  }
\label{fig:mmq10}
\end{figure}

Of the 25 waveforms presented in table~\ref{tab:Configurations}, 5 are precessing.  In figure~\ref{fig:mm}, 
for 18 out of the remaining 20 non-precessing configurations, we show the unfaithfulness between the numerical
waveforms and the analytical SEOBNRv1 waveforms as functions of the total mass.
We show the unfaithfulness of Case 9 (R10) in figure~\ref{fig:mmq10} and omit Case 11
(L3+60+00), due to its large numerical errors (see figure~\ref{fig:ErrorEstimates} and table~\ref{tab:ErrorEstimates}). 

The SEOBNRv1 model is a non-precessing spin model. It was calibrated
in~\cite{Taracchini:2012} to five non-spinning numerical waveforms with
mass ratios 1, 2, 3, 4 and 6~\cite{Buchman:2012dw} and two equal-mass spinning,
aligned/antialigned numerical waveforms (the contributed waveforms 
Cases 16 (S1+44+44) and 17 (S1$-44$$-44$)). The SEOBNRv1 model was
extrapolated to any mass ratio and to dimensionless spin values $\chi$
in the range $-1$ to $0.7$, using inspiral, merger and ringdown waveforms
produced with a Teukolsky-equation code in the large mass-ratio
limit~\cite{Barausse:2011kb}.

By definition of the unfaithfulness, the binary parameters of the signal and the template 
coincide. So, when comparing SEOBNRv1 and numerical-relativity 
waveforms, we evaluate the SEOBNRv1 waveforms using the binary parameters
extracted from the numerical simulations immediately after the junk
radiation. Whilst these parameters are somewhat gauge dependent, and we have not rigorously
included error estimates for them, the good agreement with EOB shown below indicates that this
is not a major source of error.
For clarity, we divide the results into two panels (see figure~\ref{fig:mm}). 
In each panel, we also show the numerical errors, which we compute as the unfaithfulness 
$\delta{\bar{\cal F}}$ between a numerical
waveform $h_{\rm NR}=A\rme^{\rmi\phi}$ and the waveform $h_{\text{NR, }\Delta}
= (A \pm \Delta A)\rme^{\rmi(\phi \pm \Delta\phi)}$, where $\Delta A$ and
$\Delta\phi$ are the numerical amplitude and phase errors estimated
in Sec.~\ref{sec:erroranalysis}.

As can be seen in figure~\ref{fig:mm}, at the low-mass end, in all 18 Cases, the
unfaithfulness is at most a few percent, with slopes always positive,  
indicating increasingly better agreement with the numerical-relativity
waveforms at low frequency during the last stages of inspiral. 
At the high-mass end, in 16 out of 18 cases the unfaithfulness
is below $1\%$, indicating a highly faithful modeling of 
the merger-ringdown phase in the SEOBNRv1 model. 

In Cases 8 and especially 4, the unfaithfulness is much larger than that of the 
other cases, especially when the total mass is high. We find that in these cases, the 
amplitudes and frequencies of the SEOBNRv1 waveforms have artificial 
oscillations around merger. Although these oscillations appear tiny in the time domain, they become 
much more significant in the frequency domain and lead to the substantial increase 
of unfaithfulness. It is interesting that this study with the NRAR 
waveforms has uncovered those spurious features in the EOB model that will be 
corrected in future modeling. We recall that the SEOBNRv1 model was calibrated to five non-spinning numerical 
simulations ($1 \le q \le 6$) and only two spinning non-precessing simulations with 
equal masses and equal, mild ($|\chi| = 0.44$) spin components. The  
input values~\cite{Taracchini:2012} that are extracted from the numerical waveforms and used to re-shape the EOB 
waveform around merger were extrapolated across the entire parameter space using only 
a handful of numerical simulations. Thus, it is not surprising that when we 
extrapolate the SEOBNRv1 waveforms to unequal-mass and/or unequal-spin 
configurations that are quite far from the calibrated points, we can observe some 
artefacts around merger. Although the unfaithfulness is large in these 
two cases (up to $30\%$ in Case 4), the ineffectualness is much smaller. 
We find that by minimizing only over the binary 
component masses the ineffectualness can already be reduced to below $2\%$. 
We expect a further reduction when minimizing also over the black-hole spins.
Since we estimate that the ineffectualness in all 18 cases is below $1\%$, 
we conclude that for total masses $ \geq 100M_\odot$ the SEOBNRv1 waveforms are 
sufficiently accurate for detection purposes, i.e.\ the modeling error will cause 
a loss in event rates smaller than $3\%$. 

Five cases in table \ref{tab:Configurations} (viz.\ 1, 3, 21, 22 and
23), are spinning, precessing waveforms and are not included in
figure~\ref{fig:mm}.  Since the mass ratios for these cases are not very large, 
$q=3$ for Case 21 and $q=1$ for the other four, precession-induced modulations 
are quite mild. Nevertheless, it is interesting to check whether the non-precessing SEOBNRv1 
model is able to match mildly-precessing numerical waveforms. To compare the waveforms  
in the presence of precession, at the initial time of the numerical simulation we set 
the EOB spin components in the orbital plane to zero and we identify the numerical and EOB spin
components along the direction perpendicular to the orbital 
plane. Using non-precessing waveforms generated with these
EOB spins, we find that the unfaithfulness in all five cases is a few percent at
low total masses and below $1\%$ at high total masses. The non-precessing SEOBNRv1
waveforms are therefore good approximations to the mildly precessing
numerical waveforms produced in this paper. It is worth noting that in Cases 22 and 23, the 
unfaithfulness has a negative slope toward low total masses. This 
is not surprising, since we expect that relatively stronger precessional effects 
accumulate during the inspiral, and such effects are not included in the non-precessing SEOBNRv1 
model. It is worth pointing out that these conclusions for the precessing NRAR waveforms  
refer only to the $(2,2)$ mode and only to an optimally-oriented precessing system. However, 
because the numerical waveforms are mildly precessing, we do not expect that these 
conclusions will change dramatically for a more general orientation.

In summary, for the EOB-spinning comparisons, the non-precessing
SEOBNRv1 model performs quite well outside the parameter range in which
it was originally calibrated and is also able to match mildly
precessing waveforms well. It is certainly effectual in matching all the
numerical waveforms of table~\ref{tab:Configurations}. The results
in figure~\ref{fig:mm} and the discussion above also suggest that the
faithfulness of the EOB model can be improved in the high mass-ratio, large asymmetric 
spins configurations by re-calibrating the SEOBNRv1 model using the numerical waveforms produced in this paper,
primarily Cases 4 and 8. Moreover, the 5 precessing numerical waveforms
produced in this paper can be used in the calibration of precessing EOB models.

In figure~\ref{fig:mmq10} we focus on the largest mass-ratio
numerical simulation of table \ref{tab:Configurations} (Case 9 (R10)),
which we compare with two non-spinning EOB models, since it is
non-spinning.  We use the non-spinning
limit of the SEOBNRv1 model of~\cite{Taracchini:2012,Barausse:2011kb},
which we have discussed above; the EOBNRv2 model developed
in~\cite{Pan:2011gk}, which was calibrated to five non-spinning
numerical waveforms with mass ratios 1, 2, 3, 4 and 6
from~\cite{Buchman:2012dw}, and the recent IHES-EOB model
of~\cite{Damour:2012ky}, which was also calibrated to the numerical
simulations of~\cite{Buchman:2012dw}, but incorporates the most
recent, analytical-modeling results from the
literature~\cite{Fujita:2010xj,LeTiec:2011ab,
  Barausse:2011dq,LeTiec:2011dp,Akcay:2012ea,Bini:2012ji,Bernuzzi:2012ku}
that were not available at the time the EOBNRv2 model was developed.

Case 9 (R10) is quite interesting because it is the largest
mass-ratio configuration, $q=10$, with sufficient length and accuracy 
suitable for waveform modeling, and thus it provides an
important test of the accuracy of existing EOB models that were
calibrated to comparable-mass numerical simulations of $q\leq
6$~\cite{Pan:2011gk,Taracchini:2012,Damour:2012ky} and large mass-ratio
simulations of $q \geq 1000$~\cite{Barausse:2011kb,Bernuzzi:2012ku}.
As we can see in figure~\ref{fig:mmq10}, for all three EOB models,
the unfaithfulness is a fraction of a percent. We have found that the
somewhat larger unfaithfulness of the EOBNRv2 waveform ($\sim10^{-3}$)
is due to the fact that in the EOBNRv2 model the peak of $|h_{22}|$
is enforced to be at the same time as the peak in the 
orbital frequency. By contrast, in the most recent SEOBNRv1 model, which 
includes insights from the large mass-ratio limit~\cite{Barausse:2011dq}, 
the peak of $|h_{22}|$ is chosen to occur $2.5M$ earlier 
than the peak in the orbital frequency. Nevertheless, all unfaithfulness 
is comparable to or less than the numerical error. Thus, we conclude that all the 
non-spinning EOB models perform very well, being consistent with the $q=10$ 
non-spinning numerical simulation. However, longer and more accurate numerical simulations 
are needed to improve the EOB modeling for mass ratios $q \geq 10$.

\begin{figure}
\includegraphics[width=\textwidth, bb=20 17 700 340]{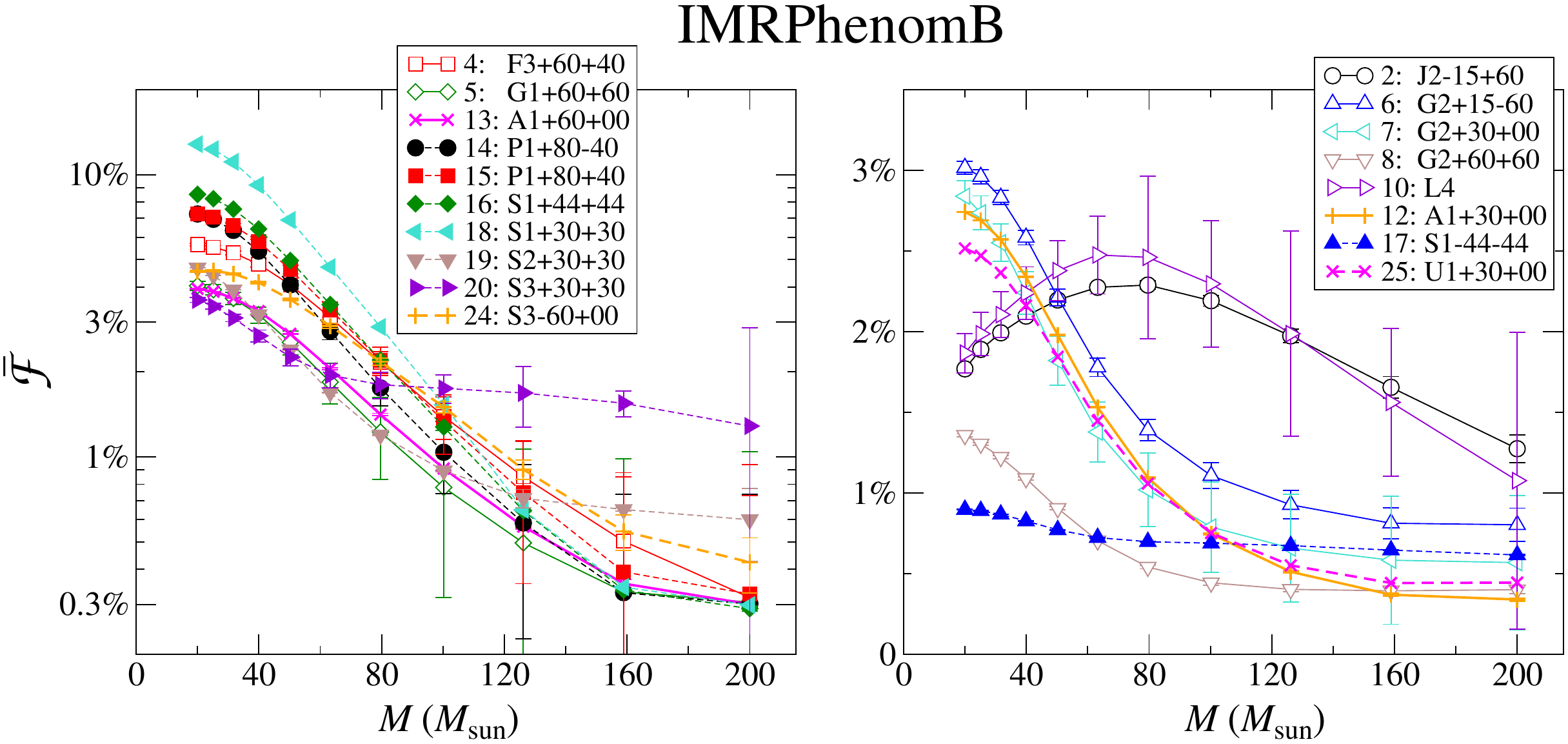}
\caption{
Unfaithfulness $\bar{\cal F}$ of the
IMRPhenomB waveform model compared to the numerical simulations.  The
horizontal axis shows the total mass, the vertical axis $\bar{\cal F}$ in percent.  The data is split into two panels for readability.
As in figure~\ref{fig:mm}, the overlaps for $M\lesssim 100M_\odot$ are 
affected by the truncation of the overlap integral at the starting frequency of the numerical 
waveforms. Numerical relativity error-estimates $\delta{\bar{\cal F}}$ on the computed unfaithfulness values are plotted, except for Cases 18 and 25 for which $\delta{\bar{\cal F}} > 20\%$.
}
\label{fig:mmphenomb}
\end{figure}
%


\subsection{Results using phenomenological waveform models}
\label{sec:Phenom}

\begin{figure}
\centering
\includegraphics[width=\textwidth]{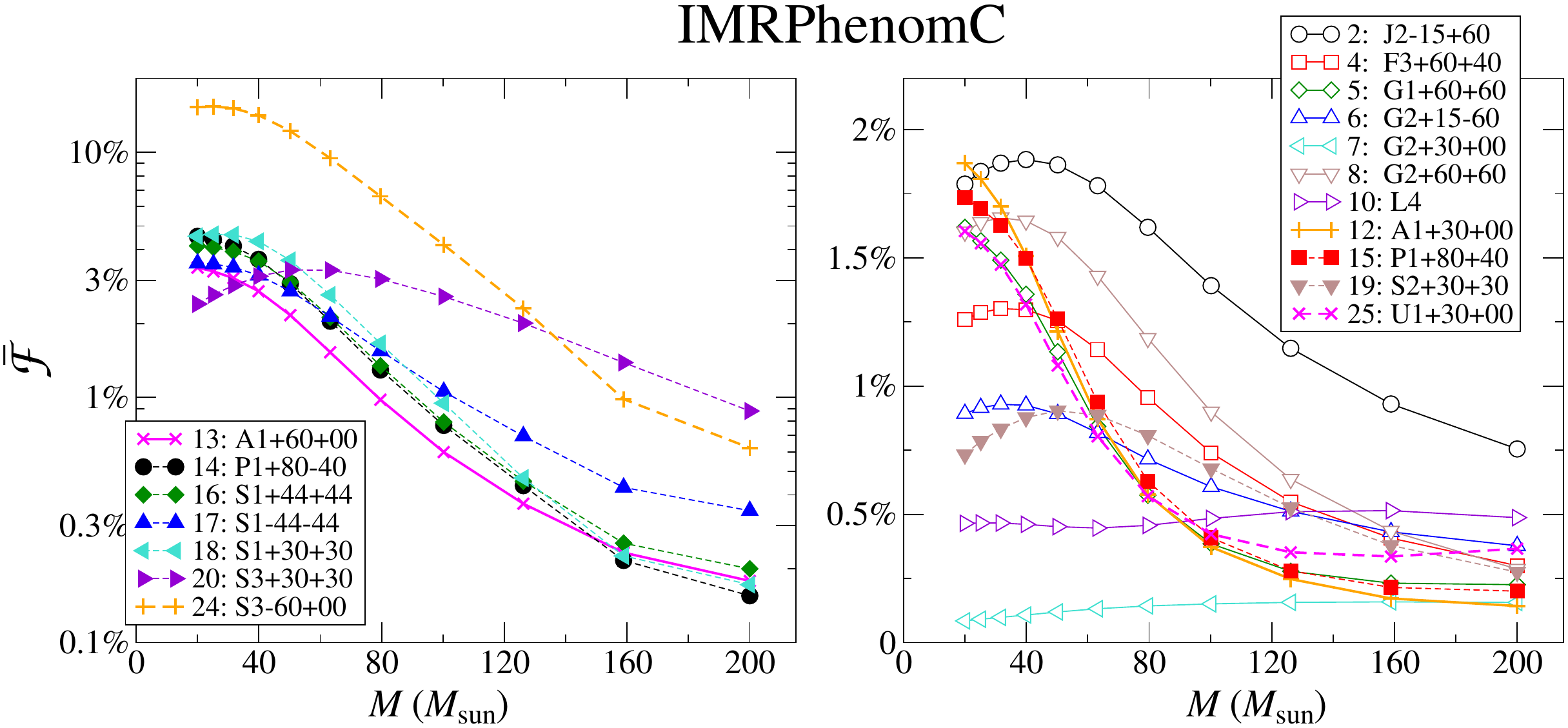}
\caption{Unfaithfulness $\bar{\cal F}$ of the IMRPhenomC waveforms
compared to the numerical simulations as functions of the total
mass. The left panel shows the cases with $\max \bar{\cal F}> 0.02$, the right panel those with $\max \bar{\cal F}< 0.02$.
As in figure~\ref{fig:mm}, the overlaps for $M\lesssim 100M_\odot$ are 
affected by the truncation of the overlap integral at the starting frequency of the numerical 
waveforms.
}

\label{fig:mmphenomc}
\end{figure}

In Figs.~\ref{fig:mmphenomb} and \ref{fig:mmphenomc}, we show the unfaithfulness of the analytical
IMRPhenomB and IMRPhenomC waveform models by comparing them with the numerical-relativity
waveforms described earlier and included in figure~\ref{fig:mm}.

The IMRPhenomB~\cite{Ajith:2011} and IMRPhenomC~\cite{Santamaria:2010yb} models are phenomenological
inspiral-merger-ringdown waveform families for non-precessing spin binaries. Both models are constructed 
in the frequency domain with somewhat different ingredients in the low frequency part. 
The waveforms are parametrized by the total mass $M$, symmetric mass ratio 
$m_1\,m_2/(m_1+m_2)^2$ and an effective spin parameter $\chi \equiv (m_1
\chi_1 + m_2 \chi_2)/M$. Since the waveforms are described by \emph{one}
spin parameter, these are not meant to be \emph{faithful} in the case
of binaries with unequal spins, and only intended as an \emph{effectual}
template family. 

IMRPhenomB is constructed by matching numerical waveforms with adiabatic PN waveforms,
notably the `TaylorT1' approximant, in a matching window. The
numerical-relativity--PN hybrid waveforms are then parametrized
in the frequency domain to get a closed form expression. The IMRPhenomC amplitude is 
constructed from a PN inspiral amplitude and a ringdown portion, 
both of which are fit to the model hybrids. IMRPhenomC uses the complete 
TaylorF2~\cite{Sathyaprakash:1991mt,Cutler:1994ys,Droz:1999qx,Ajith:2012az} PN inspiral phasing. 
Only the late inspiral and merger phases are fitted in a narrow frequency range 
$[0.1 f_\text{RD}, f_\text{RD}]$ to numerical simulations (where $f_\text{RD}$ is the ringdown frequency), 
while the ringdown waveform is obtained from analytically derived quasi-normal mode expressions 
for the frequency and attached continuously to the merger phase.

To be consistent with the unfaithfulness calculations shown in Sec.~\ref{sec:EOB}
for SEOBNRv1, we have computed the unfaithfulness for the phenomenological models as
follows. While IMRPhenomB and IMRPhenomC are naturally defined in the
Fourier domain as a power series in the frequency $f$, we have chosen
to inverse discrete Fourier transform (DFT) them so that the analysis starts out in the
time domain. The numerical-relativity and the model waveforms are
aligned at an average frequency which is computed over an interval of width
$300 M$ starting immediately after the junk radiation has passed. 
As was done for the SEOBNRv1 unfaithfulness calculation, we apply Planck tapering windows (of width
$200M$ and $20M$, respectively) to the beginning and end of both timeseries. In addition,
the waveforms are further padded with zeros before computing the DFT
to increase the frequency resolution. The low-frequency cutoff for
the unfaithfulness integral is chosen as the average frequency defined
above, which is close to the starting frequency of each numerical waveform. 
The same caveats concerning the interpretation of these
unfaithfulness results at low masses discussed at the beginning of Sec.~\ref{sec:NRARcomparison}
apply here. In addition, we should note that calculating the unfaithfulness 
directly with the frequency-domain representation
of the phenomenological models leads in general to slightly
different results. In the approach pursued here, both the
numerical-relativity and the model waveforms suffer from similar
artefacts due to the DFT near the initial frequency (and the
phenomenological waveform further exhibits errors due to the
initial inverse DFT bringing it to the time domain), while the amplitude
and phase of the phenomenological waveforms vary smoothly with
frequency if the frequency-domain representation is used.

The IMRPhenomB and IMRPhenomC models have been calibrated to waveforms of mass ratio $q \leq 4$. 
Therefore, it is not surprising that both models do not perform very well for the $q=10$ waveform 
(Case 9 (R10)), as shown in figure~\ref{fig:mmq10}. However, note that for total mass $\sim 200 M_\odot$, the 
IMRPhenomC waveform has an unfaithfulness of $3\%$. All models have an ineffectualness of $1\%$ 
for binary black holes of mass $100\mbox{--}200 M_\odot$, causing a loss of event rate of $3\%$ if used 
in advanced LIGO and Virgo searches. 

\subsection{Summary of comparison results with analytical models}
\label{sec:compare}

We find reasonable agreement between the numerical waveforms produced in 
this paper and analytical waveforms generated by the
EOBNRv2, IHES-EOB, SEOBNRv1, IMRPhenomB and IMRPhenomC models. Since the 
analytical models were not calibrated using any of the numerical waveforms 
produced by the NRAR collaboration, these results substantiate 
the ideas and procedures used in developing the above analytical models.

More quantitatively, the EOBNRv2, IHES-EOB, SEOBNRv1 waveforms agree quite well 
with the numerical waveforms except for Cases 4 and 8, where artificial features in the SEOBNRv1
waveforms around merger cause large unfaithfulness. Moreover, SEOBNRv1
waveforms are faithful toward both low and high total masses and the
unfaithfulness shows less dependence on the total mass, indicating a
consistent agreement during the inspiral, plunge, merger and ringodwn 
stages across the entire frequency range. We estimated that EOB models 
have ineffectualness below $1\%$ against all the non-precessing numerical 
waveforms produced in this paper for binary black holes with mass $100\mbox{--}200 M_\odot$, 
for which the numerical-relativity waveforms fully span the advanced LIGO bandwidth. 

The IMRPhenomB and IMRPhenomC waveforms agree quite well with numerical
waveforms at high total masses and can be quite faithful when
$M>100M_\odot$. However, the agreement is less satisfactory at low
total masses, but this is not so surprising because those waveforms use 
a single, effective spin and were developed as effectual templates. 
The unfaithfulness of most IMRPhenomB waveforms increases toward
lower total mass (with negative slope) because the phasing coefficients in 
the IMRPhenomB model do not reduce to the PN ones, but are obtained through a fit of 
the hybrid waveform. Using the TaylorF2 PN phasing, the IMRPhenomC
waveforms show improved agreement with numerical waveforms at low total 
masses. We estimated that phenomenological models have ineffectualness 
below $1\%$ against all the non-precessing numerical waveforms produced here 
for binary black holes with mass $100\mbox{--}200 M_\odot$.


\section{Conclusions}
\label{sec:concl}

The Numerical-Relativity--Analytical-Relativity (NRAR) Collaboration
has the goals of producing numerical waveforms from compact binaries and using them to
construct and validate analytical waveform models to search for binary black holes with
gravitational-wave detectors.  In this first stage of the project, 22
new waveforms were computed and rigorously analysed together with
3 contributed waveforms. The new waveforms were of higher quality
than most of the previously published waveforms.  Furthermore, a
larger region of parameter space was covered than before, including
several unequal-mass, unequal-spin binaries with aligned
spins, and several precessing runs to begin exploration of waveform
models for precessing configurations. 

For the first time, we compared analytical models previously calibrated to numerical 
waveforms to newly produced waveforms. We found that the spinning, non-precessing 
EOB model (SEOBNRv1)~\cite{Taracchini:2012}, which was previously calibrated 
to {\it only} 2 equal-mass, equal-spin, non-precessing numerical waveforms and 
5 non-spinning waveforms 
with mass ratios $q=1,2,3,4,6$ has ineffectualness below $1\%$ in matching 
all the non-precessing numerical waveforms when the binary's total mass is
$\sim 100\mbox{--}200 M_\odot$. Moreover, the non-spinning EOB models~\cite{Pan:2011gk,Taracchini:2012,Damour:2012ky} (EOBNRv2, SEOBNRv1, 
IHES-EOB) previously calibrated only to the 5 non-spinning waveforms with mass ratios $q=1,2,3,4,6$ 
have unfaithfulness below $0.3\%$ in matching the numerical waveform with mass 
ratio $q=10$. We recall that an ineffectualness of $1\%$ causes a loss of event rates 
of $\sim 3\%$. Quite interestingly, we found that the EOB unfaithfulness and ineffectualness 
is small towards both low and high total masses (except for two cases where a non-monotonicity 
of the frequency evolution close to merger causes larger unfaithfulness). This nice 
performance indicates a consistent agreement between EOB and numerical waveforms during the 
inspiral, plunge, merger and ringdown stages. These results imply that the 
EOB approach of building analytical templates is quite robust (for both 
detection and parameter estimation) in interpolating 
numerical waveforms away from the points in the parameter space where they were employed 
to calibrate the EOB templates. This success relies also on the fact that the EOB model incorporates 
accurate results from PN theory and perturbation theory, notably the Regge-Wheeler--equation and 
Teukolsky-equation results~\cite{Barausse:2011kb,Bernuzzi:2012ku}. 

We found that the phenomenological models~\cite{Ajith:2011,Santamaria:2010yb} 
(IMRPhenomB and IMPRPhenomC), which were previously calibrated to 25 spinning, non-precessing 
waveforms with mild spins and mass ratios $q \leq 4$, have somewhat larger unfaithfulness, 
especially toward low total masses. However, this is not surprising, since those models use a single, effective  
spin and were conceived as effectual templates. More importantly, their ineffectualness is below $1\%$ 
against all the non-precessing numerical waveforms produced in this paper 
for binary black holes with mass $100\mbox{--} 200 M_\odot$, for which the numerical-relativity 
waveforms fully span the advanced LIGO bandwidth. The only exception is the ineffectualness
of IMRPhenomB against the $q=10$ waveform at high masses. However, this problem does not exist in the
later,  enhanced IMRPhenomC model. The faithfulness of EOB and phenomenological models can be 
improved in the future by re-calibrating those models to the NRAR waveforms. We also 
identified a few shortcomings in both classes of waveform family, and expect that these 
can be rectified in future analytic models, using the numerical-relativity waveforms presented here.

The simulations performed pushed all participating NR groups into unknown
territory, along one or more of these dimensions:
\begin{enumerate}
 \item Simulations of more extreme parameters, for instance very
high mass-ratios --- up to $q=10$ --- and performing simulations with
both non-zero spin {\em and} unequal masses.  
\item Longer and more accurate simulations than before (20 {\em
    usable} gravitational-wave cycles before merger, with cumulative
  phase-error $\le0.25$~radians and amplitude error of $\le 1\%$).
\item Determining initial data parameters resulting in low eccentricity,
  $e\lesssim 2\times 10^{-3}$.
\item Automating codes to increase the number of simulations that can
  be performed with the available human resources.
\item Exhaustive cross-comparisons, and the resultant error checking
  and searching for the origins of discrepancies.
\end{enumerate}

In line with expectations, the accurate simulation of systems with
high spins, high mass ratios and (for finite difference codes) large
numbers of orbits proved challenging, requiring significant computational resources.
Gravitational-wave extraction required significant attention.
To push errors into the accuracy regime chosen for this project, extraction must be
performed at distances of $r > 100M$, or a more 
advanced extraction method such as CCE must be employed.
Moreover, gravitational-wave extrapolation to infinite extraction radius
(for codes that do not extract the waveforms at future null infinity)
is essential to achieve the accuracy goals. Errors at finite
extraction radius decay so slowly with increased extraction radius
that it is impractical to achieve higher gravitational-wave accuracy by simply
increasing the extraction radius.  Moreover, reliable extrapolation
requires several ($\gtrsim 5$) extraction radii which
are spaced over a significant range of radii; the ratio of outermost
extraction radius to innermost extraction radius should be $\gtrsim 3$.  
Furthermore, this project saw the first attempt to combine
  gravitational-wave extrapolation with fixed-frequency integration
 to compute $h$ from $\Psi_4$.  Unfortunately, robust error estimates are
difficult to obtain.  Future work should address this.

Following on from the NINJA project, this project represents one of
the first attempts at large-scale collaboration of this type in this
field, and it is only now becoming clear what aspects of data
management need to be improved and developed as best-practices.
The size of the data set being computed made it difficult to
collect, distribute and analyze the data across three continents.  Version control of the data
and metadata proved essential.  While the {\em Subversion} version-control system
might be adequate for the
metadata, it proved deficient to handle a data set of about 150~GByte. At the very least, metadata needs to be
provided independently of the main data set, to allow users to 
download only the small metadata files for preliminary consistency checks.
A significant improvement in data-management process for numerical relativity
would result if metadata
submissions were automatically checked and validated, allowing the
submitter to see immediately the problems with the submission, rather
than waiting for these quantities to be needed by the analysis, and
introducing a turnaround time on the order of days in each case.  

The NRAR collaboration now has the ability to take numerical-relativity waveforms in a
standard format \cite{Ajith:2007jx} and automatically process them
into a form suitable for comparison with analytic models.  While
certain consistency checks are applied automatically, this
`pipeline' can still not be used blindly, as there are frequently
complications which arise from dealing with a diverse set of waveforms
with varying characteristics. However, the existence of automated tools
makes it possible in principle to greatly expand the number of
waveforms which can be analysed as they become available.
Given how surprisingly hard this project turned out to be, one should
be mindful that further improvements to numerical simulations might also be
hard, e.g.\ doubling the time span, halving errors, or going to
untested regions of parameter space.  While numerical relativity has
made tremendous progress over the last years, pushing beyond the
current limits of simulations is still a research project where
unpredictable new problems can and do arise.

\vspace{0.5truecm}

We note that during the last stages of completion of this manuscript two papers appeared
on arXiv reporting the production of hundreds of new numerical-relativity 
waveforms~\cite{Pekowsky:2013ska,Mroue:2013xna}. The two papers were written by two 
of the NR groups that are part of the NRAR collaboration. When planning the next steps of 
the NRAR collaboration, these results will be taken into account to avoid duplication of 
numerical simulations.


\ack



We thank Beverly Berger and Kip Thorne for helping to establish the NRAR collaboration, 
Duncan Brown and Frans Pretorius for their work as members of the 
NRAR executive committee, Duncan Brown, Evan Ochsner and B. Sathyaprakash 
for useful discussions in the first stage of the NRAR collaboration when 
the scientific plan and accuracy requirements were established. We also 
thank John Baker, Peter Diener, Bernard Kelly and Riccardo Sturani for 
useful interactions. It is also a pleasure to thank David Hilditch for useful 
discussions on the hyperbolicity of BSSN, Roman Gold for useful discussions on
requirements for accurate computation of gravitational waves and
P.~Ajith for help with performing and interpreting the IMRPhenom
comparisons.  We also thank Bruce
Loftis and Christal Yost at the National Institute for Computational
Sciences (NICS) for their support.


Caltech acknowledges support from the Sherman Fairchild Foundation
and NSF grants
PHY-1068881, PHY-1005655, and DMS-1065438.
M. Hannam was supported by Science and Technology Facilities Council 
grants ST/H008438/1 and ST/I001085/1.
CITA acknowledges support from NSERC of Canada, the Canada Chairs
Program, and the Canadian Institute for Advanced Research.
Cornell acknowledges support from the Sherman Fairchild Foundation
and NSF grants PHY-0969111 and PHY-1005426.
FAU acknowledges financial support via the NSF grants 0855315 and 1204334. 
GA Tech acknowledges financial support via NSF grants 1205864,
1212433, 0903973, 0941417, and 0955825.
Jena and the AEI acknowledge financial support via the DFG
SFB/Transregio 7.
U.~Sperhake acknowledges financial support via the Ram{\'on} y Cajal
Programme of the Ministry of Education and Science in Spain, the
FP7-PEOPLE-2011-CIG grant 293412 CBHEO, the STFC GR Roller grant
ST/I002006/1, the ERC Starting grant ERC-2010-StG DyBHo, and the
FP7-PEOPLE-2011-IRSES grant 295189 NRHEP\@. A.~Nerozzi acknowledges
support by the Funda\c c\~ao para a Ci\^encia e
Tecnologia through grants SFRH/BPD/47955/2008 and
PTDC/FIS/116625/2010. H.~Witek acknowledges support
by the {\it ERC-2011-StG
279363--HiDGR} ERC Starting Grant and by the {\it DyBHo--256667} ERC
Starting Grant. H.W. acknowledges support by FCT--Portugal through grant
nos. SFRH/BD/46061/2008 and CERN/FP/123593/2011 at early stages of this
work.
C.~Reisswig acknowledges financial support from NASA via the Einstein
Postdoctoral Fellowship grant PF2-130099 awarded by the Chandra X-ray
center, which is operated by the Smithsonian Astrophysical Observatory
for NASA under contract NAS8-03060.
RIT acknowledges financial support via NSF grants AST-1028087,
PHY-0929114, PHY-0969855, PHY-0903782, OCI-0832606, and DRL-1136221,
and NASA grant 07-ATFP07-0158. H.~Nakano would also like to acknowledge
support by the Grand-in-Aid for Scientific Research (24103006).
UIUC acknowledges financial support via NSF grants
PHY-0963136, AST-1002667, and NASA grants NNX11AE11G,
NNX13AH44G\@. V.~Paschalidis would also like to acknowledge 
financial support from a Fortner Research Fellowship.
UMD acknowledges financial support via NSF grants PHY-0903631 and
PHY-1208881, and NASA grants NNX09AI81G and NNX12AN10G\@.

We also acknowledge the hospitality of the Kavli Institute for
Theoretical Physics during the program ``Chirps, Mergers and
Explosions'' (supported by the NSF grant PHY11-25915).


Computing time for this project was provided by the NSF on the
TeraGrid (now XSEDE) systems Athena and Kraken at NICS\@.
AEI used computational resources on the Datura cluster at AEI\@.
Caltech and Cornell used additional computational resources on the
Zwicky cluster at Caltech, which is supported by
the Sherman Fairchild Foundation and by NSF award PHY-0960291, and
on the NSF XSEDE network under grant TG-PHY990007N\@.
CITA used additional computational resources at the SciNet HPC
Consortium on the GPC system.
FAU used additional computational resources via TeraGrid allocations
TG-AST100021 and TG-PHY100051 at NICS on the Kraken system.
GA Tech used additional computational resources via XSEDE allocation
TG-PHY120016, and at Georgia Tech on the Cygnus cluster.
JCP used additional computational resources at the LRZ, Munich, and
 as part of the European PRACE petascale computing
initiative on the clusters Hermit, Curie and SuperMUC; simulations were
 were also carried out at Advanced Research Computing 
(ARCCA) at Cardiff.
US, AN and HW used additional computational resources via XSEDE allocation
PHY-090003 at SDCS on the Trestles system and at NICS on the Kraken
system, via the Barcelona Supercomputing Center (BSC) allocation
AECT-2012-3-0011, via the Centro de Supercomputaci{\'o}n de Galicia
(CESGA) allocation ICTS-2011-234, the Baltasar-Sete-S\'ois cluster
at CENTRA/IST Lisbon,
the CANE cluster in Poland through PRACE DECI-7
and at the DiRAC HPC Facility on the
COSMOS supercomputer supported by STFC and BIS\@.
RIT used additional computational resources at TACC on the Ranger
system via XSEDE allocation TG-PHY060027N, and at RIT on the
NewHorizons/BlueSky system supported by NSF grants PHY-0722703, DMS-0820923,
AST-1028087 and PHY-1229173.
UIUC used additional computational resources at TACC on the Stampede
system via XSEDE allocation TG-MCA99S008.
Hosting for the NRAR data set and collaboration tools was provided by
the Syracuse University Gravitational-wave Group, supported by NSF
awards PHY-0847611, PHY-1040231 and PHY-1104371.

The EOBNRv2, SEOBNRv1, IMRPhenomB and IMRPhenomC templates were produced 
using the public LIGO Algorithm Library~\cite{LALwebsite}, while the IHES-EOB templates 
were produced using the public code in \cite{IHES-EOBwebsite}.


\section*{References}
\bibliographystyle{iopart-num}
\bibliography{references}


\appendix

\section{Initial Data Parameters}
\label{appendixid}

\begin{table*}[!ht]
\caption{\label{tab:InitialData1} Initial data: mass parameters and
  positions.  Suffixes $1$ and $2$ denote the two black holes. $\mu_i$ is the
  initial-data mass parameter (see code descriptions).  $M$ is the sum
  of the Christodoulou masses of each black hole at the `after-junk' time
  $t_0$. $\mathbf{x}_i$ is the coordinate position of each black hole.
  Vectors
  are given with subscripts $x$ and $y$ in a Cartesian coordinate system.  The black holes
  lie in the $xy$ plane.}
\footnotesize
\begin{flushleft}
\begin{tabular}{llllllll}
\br
   &       &           &           & \centre{2}{$\mathbf{x}_{1}/M$}  & \centre{2}{$\mathbf{x}_{2}/M$} \\
   &       &           &           & \crule{2}  & \crule{2} \\
\# & Label & $\mu_1/M$ & $\mu_2/M$ & $x$ & $y$ & $x$ & $y$ \\
\mr
$1$ & J1p+49+11 & $0.42386572$ & $0.47868401$ & $0$ & $6.2562331$ & $0$ & $-6.2562331$ \\
$2$ & J2-15+60 & $0.64865948$ & $0.26550575$ & $0$ & $-3.3333222$ & $0$ & $6.6666444$ \\
$3$ & F1p+30-30 & $0.40458816$ & $0.40467792$ & $0$ & $5.9847297$ & $0$ & $-5.9847297$ \\
$4$ & F3+60+40 & $0.61400511$ & $0.22334067$ & $0$ & $-2.6817824$ & $0$ & $8.0453472$ \\
$5$ & G1+60+60 & $0.40477256$ & $0.40477255$ & $-5.9999999$ & $0$ & $5.9999999$ & $0$ \\
$6$ & G2+15-60 & $0.65061114$ & $0.26737099$ & $4.$ & $0$ & $-8.0000001$ & $0$ \\
$7$ & G2+30+00 & $0.63208139$ & $0.32260272$ & $4.$ & $0$ & $-8.$ & $0$ \\
$8$ & G2+60+60 & $0.54353789$ & $0.26665158$ & $3.6666667$ & $0$ & $-7.3333333$ & $0$ \\
$9$ & R10 & $0.90392578$ & $0.084911211$ & $-0.75029467$ & $0$ & $7.60393$ & $0$ \\
$10$ & L4 & $0.79227685$ & $0.19127667$ & $2.1865032$ & $0$ & $-8.7460126$ & $0$ \\
$11$ & L3+60+00 & $0.61425068$ & $0.23827383$ & $-2.329625$ & $0$ & $6.988875$ & $0$ \\
$12$ & A1+30+00 & $0.4881057$ & $0.46983416$ & $5.8999302$ & $0$ & $-5.8999302$ & $0$ \\
$13$ & A1+60+00 & $0.48812903$ & $0.40453387$ & $5.8998442$ & $0$ & $-5.8998442$ & $0$ \\
$14$ & P1+80-40 & $0.45533296$ & $0.30397002$ & $6.2506584$ & $0$ & $-6.2506584$ & $0$ \\
$15$ & P1+80+40 & $0.45524795$ & $0.30394401$ & $6.1759493$ & $0$ & $-6.1759493$ & $0$ \\
$16$ & S1+44+44 & $0.49986559$ & $0.49986559$ & $6.6810546$ & $0$ & $-6.6810546$ & $0$ \\
$17$ & S1-44-44 & $0.49986355$ & $0.49986355$ & $6.6753871$ & $0$ & $-6.6753871$ & $0$ \\
$18$ & S1+30+30 & $0.49996958$ & $0.49996958$ & $7.4995437$ & $-5.0912558\times 10^{-8}$ & $-7.4995437$ & $-5.0912558\times 10^{-8}$ \\
$19$ & S2+30+30 & $0.66662601$ & $0.333313$ & $4.6301375$ & $0.01288301$ & $-9.2949462$ & $0.01288301$ \\
$20$ & S3+30+30 & $0.74995534$ & $0.24998511$ & $3.3712598$ & $0.016546877$ & $-10.171299$ & $0.016546877$ \\
$21$ & S3p+00-15 & $0.74998557$ & $0.24999519$ & $3.5532079$ & $0.016524038$ & $-10.706707$ & $0.016524038$ \\
$22$ & S1p+30+30 & $0.49996927$ & $0.49996927$ & $7.359082$ & $0$ & $-7.359082$ & $0$ \\
$23$ & S1p-30-30 & $0.50001$ & $0.50001$ & $8.0977416$ & $1.8034293\times 10^{-6}$ & $-8.0977488$ & $1.8034293\times 10^{-6}$ \\
$24$ & S3-60+00 & $0.74943027$ & $0.24981009$ & $3.4836516$ & $0.016192245$ & $-10.505713$ & $0.016192245$ \\
$25$ & U1+30+00 & $0.46984424$ & $0.4881112$ & $-5.9$ & $0$ & $5.9$ & $0$ \\

\br
\end{tabular}
\end{flushleft}
\end{table*}
\begin{table*}[!ht]
\caption{\label{tab:InitialData2} Initial data: velocities/momenta.
  Suffixes $1$ and $2$ denote the two black holes. $M$ is the sum of the
  Christodoulou masses of each black hole at the `after-junk' time $t_0$. For
  Bowen-York configurations, the momenta $\mathbf{p}_i$ are given,
  whereas for Conformal-Thin-Sandwich configurations (those with label
  initial S), the velocities $\mathbf{V}_i$ are given. Vectors are given
  as $v_x \, v_y \, v_z$ in a Cartesian coordinate system.}
\footnotesize
\begin{flushleft}
\begin{tabular}{llllllll}
\br
%
%
   &       & \centre{3}{$10^3 (\mathbf{p}_1/M | \mathbf{V}_1)$}                                          & \centre{3}{$10^3 (\mathbf{p}_2/M | \mathbf{V}_2)$} \\
   &       & \crule{3}                                          & \crule{3} \\
\# & Label & $x$ & $y$ & $z$ & $x$ & $y$ & $z$ \\
\mr
$1$ & J1p+49+11 & $-81.586324$ & $-0.44156253$ & $0.38454864$ & $81.586324$ & $0.44156253$ & $-0.38454864$ \\
$2$ & J2-15+60 & $85.163716$ & $0.76349079$ & $0$ & $-85.163716$ & $-0.76349079$ & $0$ \\
$3$ & F1p+30-30 & $-85.280216$ & $-0.38705042$ & $0$ & $85.280216$ & $0.38705042$ & $0$ \\
$4$ & F3+60+40 & $66.395903$ & $0.24967205$ & $0$ & $-66.395903$ & $-0.24967205$ & $0$ \\
$5$ & G1+60+60 & $0.47254707$ & $-82.397999$ & $0$ & $-0.47254707$ & $82.397999$ & $0$ \\
$6$ & G2+15-60 & $-0.43487079$ & $75.97791$ & $0$ & $0.43487079$ & $-75.97791$ & $0$ \\
$7$ & G2+30+00 & $-0.40590058$ & $74.816237$ & $0$ & $0.40590058$ & $-74.816237$ & $0$ \\
$8$ & G2+60+60 & $-0.49753473$ & $77.378061$ & $0$ & $0.49753473$ & $-77.378061$ & $0$ \\
$9$ & R10 & $0.16787436$ & $-36.558416$ & $0$ & $-0.16787436$ & $36.558416$ & $0$ \\
$10$ & L4 & $-0.38936512$ & $58.050509$ & $0$ & $0.38936512$ & $-58.050509$ & $0$ \\
$11$ & L3+60+00 & $0.48767064$ & $-73.103363$ & $0$ & $-0.48767064$ & $73.103363$ & $0$ \\
$12$ & A1+30+00 & $-0.51501591$ & $85.285491$ & $0$ & $0.51501591$ & $-85.285491$ & $0$ \\
$13$ & A1+60+00 & $-0.51904429$ & $84.698363$ & $0$ & $0.51904429$ & $-84.698363$ & $0$ \\
$14$ & P1+80-40 & $-0.45368779$ & $82.107448$ & $0$ & $0.45368779$ & $-82.107448$ & $0$ \\
$15$ & P1+80+40 & $-0.43446276$ & $80.958927$ & $0$ & $0.43446276$ & $-80.958927$ & $0$ \\
$16$ & S1+44+44 & $-0.33287494$ & $123.67003$ & $0$ & $0.33287494$ & $-123.67003$ & $0$ \\
$17$ & S1-44-44 & $-0.74710129$ & $125.43959$ & $0$ & $0.74710129$ & $-125.43959$ & $0$ \\
$18$ & S1+30+30 & $-0.2554492$ & $118.21853$ & $0$ & $0.2554508$ & $-118.21852$ & $0$ \\
$19$ & S2+30+30 & $-0.41767515$ & $81.031813$ & $0$ & $0.16039033$ & $-162.67201$ & $0$ \\
$20$ & S3+30+30 & $-0.42953692$ & $58.999873$ & $0$ & $0.13264816$ & $-178.00901$ & $0$ \\
$21$ & S3p+00-15 & $-0.45466925$ & $60.404889$ & $0$ & $0.24265399$ & $-182.01832$ & $0$ \\
$22$ & S1p+30+30 & $0.68885058$ & $117.0166$ & $0$ & $-0.68885058$ & $-117.0166$ & $0$ \\
$23$ & S1p-30-30 & $-0.650836$ & $115.79539$ & $0$ & $0.650785$ & $-115.79549$ & $0$ \\
$24$ & S3-60+00 & $-0.448744$ & $61.2985$ & $0$ & $0.209116$ & $-184.862$ & $0$ \\
$25$ & U1+30+00 & $0.515$ & $-85.29$ & $0$ & $-0.515$ & $85.29$ & $0$ \\

\br
\end{tabular}
\end{flushleft}
\end{table*}
\begin{table*}[!ht]
\caption{\label{tab:InitialData3} Initial data: spins.  Suffixes $1$
  and $2$ denote the two black holes. $M$ is the sum of the Christodoulou
  masses of each black hole at the `after-junk' time $t_0$. $\mathbf{S}_i$ is
  the spin angular momentum of each black hole. Vectors are given as $v_x \,
  v_y \, v_z$ in a Cartesian coordinate system.}
\footnotesize
\begin{flushleft}
\begin{tabular}{llllllll}
\br
   &       & \centre{3}{$\mathbf{S}_1/M^2$} & \centre{3}{$\mathbf{S}_2/M^2$} \\
   &       & \crule{3} & \crule{3} \\
\# & Label & $x$ & $y$ & $z$ & $x$ & $y$ & $z$ \\
\mr
$1$ & J1p+49+11 & $-0.0421608$ & $-0.032117926$ & $0.12366979$ & $0.037330398$ & $0.032234057$ & $0.026242352$ \\
$2$ & J2-15+60 & $0$ & $0$ & $-0.066666222$ & $0$ & $0$ & $0.066666222$ \\
$3$ & F1p+30-30 & $0.12990416$ & $0$ & $0.075000204$ & $0$ & $0.12990416$ & $-0.075000204$ \\
$4$ & F3+60+40 & $0$ & $0$ & $0.3375$ & $0$ & $0$ & $0.025$ \\
$5$ & G1+60+60 & $0$ & $0$ & $0.15$ & $0$ & $0$ & $0.15$ \\
$6$ & G2+15-60 & $0$ & $0$ & $0.066666667$ & $0$ & $0$ & $-0.066666667$ \\
$7$ & G2+30+00 & $0$ & $0$ & $0.13333333$ & $0$ & $0$ & $0$ \\
$8$ & G2+60+60 & $0$ & $0$ & $0.26666667$ & $0$ & $0$ & $0.066666667$ \\
$9$ & R10 & $0$ & $0$ & $0$ & $0$ & $0$ & $0$ \\
$10$ & L4 & $0$ & $0$ & $0$ & $0$ & $0$ & $0$ \\
$11$ & L3+60+00 & $0$ & $0$ & $0.3375$ & $0$ & $0$ & $0$ \\
$12$ & A1+30+00 & $0$ & $0$ & $0$ & $0$ & $0$ & $0.074998226$ \\
$13$ & A1+60+00 & $0$ & $0$ & $0$ & $0$ & $0$ & $0.14999208$ \\
$14$ & P1+80-40 & $0$ & $0$ & $-0.10002107$ & $0$ & $0$ & $0.20004214$ \\
$15$ & P1+80+40 & $0$ & $0$ & $0.10002107$ & $0$ & $0$ & $0.20004214$ \\
$16$ & S1+44+44 & $0$ & $0$ & $0.10914509$ & $0$ & $0$ & $0.10914509$ \\
$17$ & S1-44-44 & $0$ & $0$ & $-0.10940641$ & $0$ & $0$ & $-0.10940641$ \\
$18$ & S1+30+30 & $0$ & $0$ & $0.074990874$ & $0$ & $0$ & $0.074990874$ \\
$19$ & S2+30+30 & $0$ & $0$ & $0.13331707$ & $0$ & $0$ & $0.033329267$ \\
$20$ & S3+30+30 & $0$ & $0$ & $0.1687299$ & $0$ & $0$ & $0.018747767$ \\
$21$ & S3p+00-15 & $0$ & $0$ & $0$ & $0.016237351$ & $0$ & $-0.0093746393$ \\
$22$ & S1p+30+30 & $0.12988784$ & $0$ & $0.074990781$ & $0.12988784$ & $0$ & $0.074990781$ \\
$23$ & S1p-30-30 & $0$ & $0.12990901$ & $-0.075003$ & $0$ & $0.12990901$ & $-0.075003$ \\
$24$ & S3-60+00 & $0$ & $0$ & $-0.33698744$ & $0$ & $0$ & $0$ \\
$25$ & U1+30+00 & $0$ & $0$ & $0.075$ & $0$ & $0$ & $0$ \\

\br
\end{tabular}
\end{flushleft}
\end{table*}

\clearpage

\section{Waveform Error Estimates}
\label{appendixerrs}
\begin{table}[!ht]
\caption{\label{tab:ErrorEstimates}Error estimates.  For each waveform,
  the table shows the number of cycles in the $(2,2)$ mode
  of $h$ between the start of the usable NR waveform and the peak of
  the amplitude of $h$ and the relative amplitude and absolute phase
  errors measured at the start of the inspiral and at the reference
  time shortly before the merger.  The `aligned' errors were computed
  using waveforms which have been aligned as described in
  Sec.~\ref{sec:erroranalysis}.}
\footnotesize
\begin{tabular}{lllllllll}
\br
\# & Label & Cycles & $\Delta A/A|_\text{insp.}$ & $\Delta \phi|_{\text{insp.}}$ & $\Delta A/A|_{\text{ref.}}$ & $(\Delta A/A)|_{\text{ref.}}^{\text{aligned}}$ & $\Delta \phi|_{\text{ref}.}$ & $\Delta \phi|_{\text{ref}.}^{\text{aligned}}$ \\
\mr
$1$ & J1p+49+11 & $25$ & $0.023$ & $0.046$ & $0.018$ & $0.028$ & $0.16$ & $0.16$ \\
$2$ & J2-15+60 & $13$ & $0.0054$ & $0.02$ & $0.0074$ & $0.0095$ & $0.057$ & $0.016$ \\
$3$ & F1p+30-30 & $19$ & $0.15$ & $0.41$ & $0.068$ & $0.068$ & $1.3$ & $0.98$ \\
$4$ & F3+60+40 & $24$ & $0.014$ & $0.064$ & $0.018$ & $0.012$ & $0.34$ & $0.2$ \\
$5$ & G1+60+60 & $25$ & $0.014$ & $0.078$ & $0.012$ & $0.017$ & $0.29$ & $0.55$ \\
$6$ & G2+15-60 & $19$ & $0.013$ & $0.068$ & $0.04$ & $0.019$ & $0.67$ & $0.37$ \\
$7$ & G2+30+00 & $23$ & $0.015$ & $0.073$ & $0.024$ & $0.024$ & $0.43$ & $0.54$ \\
$8$ & G2+60+60 & $22$ & $0.0067$ & $0.045$ & $0.0048$ & $0.0094$ & $0.048$ & $0.086$ \\
$9$ & R10 & $13$ & $0.079$ & $0.3$ & $0.041$ & $0.019$ & $0.54$ & $0.16$ \\
$10$ & L4 & $20$ & $0.075$ & $0.07$ & $0.083$ & $0.056$ & $1.3$ & $0.53$ \\
$11$ & L3+60+00 & $20$ & $0.041$ & $0.037$ & $0.17$ & $0.09$ & $5.5$ & $2.6$ \\
$12$ & A1+30+00 & $20$ & $0.0017$ & $0.011$ & $0.015$ & $0.009$ & $0.23$ & $0.07$ \\
$13$ & A1+60+00 & $22$ & $0.0038$ & $0.0093$ & $0.015$ & $0.031$ & $0.16$ & $0.049$ \\
$14$ & P1+80-40 & $24$ & $0.0009$ & $0.0083$ & $0.035$ & $0.014$ & $0.87$ & $0.32$ \\
$15$ & P1+80+40 & $24$ & $0.0011$ & $0.0076$ & $0.025$ & $0.01$ & $0.66$ & $0.28$ \\
$16$ & S1+44+44 & $30$ & $0.0028$ & $0.02$ & $0.0041$ & $0.0054$ & $0.11$ & $0.13$ \\
$17$ & S1-44-44 & $20$ & $0.0016$ & $0.019$ & $0.0038$ & $0.0059$ & $0.1$ & $0.12$ \\
$18$ & S1+30+30 & $37$ & $0.0027$ & $0.018$ & $0.017$ & $0.012$ & $0.48$ & $0.13$ \\
$19$ & S2+30+30 & $34$ & $0.0023$ & $0.032$ & $0.0032$ & $0.0042$ & $0.022$ & $0.019$ \\
$20$ & S3+30+30 & $37$ & $0.0042$ & $0.032$ & $0.0049$ & $0.0051$ & $0.058$ & $0.04$ \\
$21$ & S3p+00-15 & $36$ & $0.0036$ & $0.019$ & $0.02$ & $0.019$ & $0.39$ & $0.21$ \\
$22$ & S1p+30+30 & $35$ & $0.012$ & $0.1$ & $0.069$ & $0.023$ & $1.3$ & $0.26$ \\
$23$ & S1p-30-30 & $34$ & $0.0058$ & $0.006$ & $0.064$ & $0.06$ & $1.2$ & $1.1$ \\
$24$ & S3-60+00 & $27$ & $0.0039$ & $0.031$ & $0.04$ & $0.047$ & $0.51$ & $0.59$ \\
$25$ & U1+30+00 & $20$ & $0.04$ & $0.073$ & $0.063$ & $0.044$ & $0.99$ & $0.82$ \\

\br
\end{tabular}
\end{table}

\end{document}